\documentclass[useAMS,usegraphicx,usenatbib]{mn2e}
\usepackage{threeparttable}
\usepackage{times}
\usepackage{epsfig}
\usepackage{textcomp}
\usepackage{amssymb}
\usepackage{subfigure}
\usepackage{multirow}
\usepackage{enumerate}
\usepackage{amsmath}
\usepackage{pdflscape}
\usepackage{multirow}
\usepackage{graphicx}
\usepackage{appendix}

\newcommand{\photozs}{photo-$z$'s}
\newcommand{\spitzer}{\textit{Spitzer}}
\newcommand{\lir}{L$_\mathrm{IR}$}
\newcommand{\nulseventy}{L$_{70}$}
\newcommand{\nulonesixty}{L$_{160}$}
\newcommand{\loglonesixty}{$\log{L_{160}}$}
\newcommand{\lseventy}{L$_{70}$}
\newcommand{\lonesixty}{L$_{160}$}
\newcommand{\loglseventy}{$\log~L_{70}$}
\newcommand{\nulnu}{$\nu$L$_{\nu}$}
\newcommand{\vmax}{1/$V_\mathrm{max}$}
\newcommand{\alphal}{\hbox{$\alpha_\textsc{l}$}}
\newcommand{\alphad}{$\alpha_\textsc{d}$}

\newcommand{\loglstar}{$\log L^\ast$}
\newcommand{\logphistar}{$\log \phi^\ast$}
\newcommand{\loglir}{$\log$ L$_\mathrm{IR}$}
\newcommand{\lsun}{L$_\odot$}

\newcommand{\omegair}{$\Omega_\mathrm{IR}$}
\newcommand{\sfrunits}{M$_\odot$ yr$^{-1}$ Mpc$^{-3}$}
\newcommand{\rhosfr}{$\rho_\textsc{sfr}$}
\newcommand{\iras}{\textit{IRAS}}
\newcommand{\iso}{\textit{ISO}}
\newcommand{\herschel}{\textit{Herschel}}
\newcommand{\akari}{\textit{AKARI}}

\newcommand{\sqrdeg}{deg$^{2}$}
\newcommand{\mpcdex}{Mpc$^{-3}$}

\title[Evolution of far-infrared luminosity functions in SWIRE]{Evolution of the far-infrared luminosity functions in the \textit{Spitzer} Wide-area Infrared Extragalactic Legacy Survey}

\author[H. Patel et~al.]{H.~Patel,$^1$$\thanks{harsit.patel08@imperial.ac.uk}$ D.~L.~Clements,$^1$ M.~Vaccari,$^2$ D.~J.~Mortlock,$^{1,3}$ M.~Rowan-Robinson,$^1$
\newauthor~I.~P{\'e}rez-Fournon,$^4$\\
$^1$Astrophysics Group, Imperial College London, Blackett Laboratory, Prince Consort Road, London, SW7~2AW\\
$^2$Dipartimento di Astronomia, Universit\`{a} di Padova, vicolo Osservatorio, 3, 35122 Padova, Italy\\
$^3$Department of Mathematics, Imperial College London, Prince Consort Road, London, SW7~2AW\\
$^4$Instituto de Astrof\'{i}sica de Canarias, C/ V\'{i}a L\'{a}ctea s/n, E-38200 La Laguna, Spain}

\begin{document}

\date{Accepted. Received}

\pagerange{\pageref{firstpage}--\pageref{lastpage}} \pubyear{2011}

\maketitle

\label{firstpage}

\begin{abstract}
We present new observational determination of the evolution of the rest-frame 70 and 160 {\micron} and total infrared (TIR) galaxy luminosity functions (LFs) using 70 {\micron} data from the \textit{Spitzer} Wide-area Infrared Extragalactic Legacy Survey (SWIRE). The LFs were constructed for sources with spectroscopic redshifts only in the XMM-LSS and Lockman Hole fields from the SWIRE photometric redshift catalogue. The 70 {\micron} and TIR LFs were constructed in the redshift range $0<z<1.2$ and the 160 {\micron} LF was constructed in the redshift range $0<z<0.5$ using a parametric Bayesian and the {\vmax} methods. We assume in our models, that the faint-end power-law index of the LF does not evolve with redshifts. We find the the double power-law model is a better representation of the IR LF than the more commonly used power-law and Gaussian model. We model the evolution of the FIR LFs as a function of redshift where where the characteristic luminosity, $L^\ast$ and the LF normalisation, $\phi^\ast$, evolve as $\propto(1+z)^{\alpha_\textsc{l}}$ and $\propto(1+z)^{\alpha_\textsc{d}}$ respectively. The rest-frame 70 {\micron} LF shows a strong luminosity evolution and negligible density evolution out to $z=1.2$ with {\alphal}$=3.44^{+0.20}_{-0.18}$. The rest-frame 160 {\micron} LF also showed rapid luminosity evolution with {\alphal}$=3.70^{+0.18}_{-0.24}$ and {\alphad}$=-0.04^{+0.57}_{-0.40}$ out to $z=0.5$.  The rate of evolution in luminosity and density is consistent with values estimated from previous studies using data from {\iras}, {\iso} and {\spitzer}. The TIR LF evolves in luminosity with {\alphal}$=3.53^{+0.20}_{-0.21}$ and the normalisation evolves as {\alphad}$=-0.15^{+0.39}_{-0.34}$ which is in agreement with previous results from {\spitzer} 24 {\micron} which find strong luminosity evolution and marginal density evolution. By integrating the LF we calculated the co-moving IR luminosity density out to $z=1.2$, which confirm the rapid evolution in number density of LIRGs which contribute $\sim70\%$ to the co-moving star formation rate density at $z=1.2$. Furthermore, our model suggest that by $z=1.2$, ULIRGs are responsible for $\sim10\%$ of the IR luminosity density. Our results based on 70 {\micron} data confirms that the bulk of the star formation at $z=1$ takes place in dust obscured objects.

\end{abstract}

\begin{keywords}
Galaxies: evolution -- infrared: galaxies -- galaxies: starburst -- cosmology: observations
\end{keywords}

\section{INTRODUCTION}
\label{sect:introswirelf}
The galaxy luminosity function (LF) and its evolution is an important probe which describes the distribution of galaxies as a function of luminosity over the history of the Universe. LFs have been used to constrain galaxy formation and evolution models and to quantify the evolution of the star formation rate (SFR). The infrared (IR) LF is essential to understand the amount of energy released by reprocessed dust emission from star formation and active galactic nuclei (AGN). The discovery of IR luminous galaxies from ground based photometry \citep{rieke197217695} and by the \textit{Infrared Astronomical Satellite} ({\iras}) have found them to be locally very rare and to only contribute $\sim$ 5\% to the local IR luminosity density \citep{soifer1991101354}. The detection of the Cosmic Infrared Background (CIB; \citealt{puget19963085, fixsen1998508123}) however, has shown that roughly half of the energy released in the Universe has been absorbed by dust and re-radiated into the IR, which implies that dust obscured star formation was much more important at higher redshifts.

\indent Studies using observations performed by {\iras} and the \textit{Infrared Space Observatory} ({\iso}) have shown that dusty star forming galaxies have undergone strong evolution, as demonstrated by their LFs. \cite{saunders1990242318} construct the 60 {\micron} and $40-120$ {\micron} far-IR (FIR) LFs based on {\iras} observations finding strong luminosity evolution (modelled as $L^\ast(z)\propto (1+z)^\text{\alphal}$, where $L^\ast$ is the characteristic luminosity and $z$ is the redshift) with {\alphal} = $3\pm1$.  Similar rates of positive evolution ({\alphal} = $3 - 5$) are seen in LFs constructed from ISO surveys at 12 {\micron} \citep{clements2001325665}, 90 {\micron} \citep{serjeant2004355813} and 170 {\micron} \citep{takeuchi2006448525}. \cite{pozzi2004609122} determine the 15 {\micron} LF of galaxies from the European Large Area {\iso} Survey (ELAIS) to find that the starburst population evolves both in luminosity, with {\alphal} = 3.5, and density (modelled as $\phi^\ast(z) \propto (1+z)^\text{\alphad}$, where $\phi^\ast$ is the characteristic number density) with {\alphad} = 3.8 being consistent with model predictions of source counts and redshift distribution.

\indent The sensitivity and spatial resolution of the \textit{Spitzer Space Telescope} (\textit{Spitzer}; \citealt{werner20041541}) has revolutionised our understanding of the evolution of IR galaxies particularly at high redshifts ($z > 1$).  Several studies based on Multiband Imager for {\spitzer} (MIPS) 24 {\micron} observations have been used to construct the rest-frame 8 {\micron} \citep{babbedge20063701159, caputi200766097, rodighiero20105158}, 12 {\micron} \citep{perezgonzalez200563082}, 15{\micron} \citep{lefloch2005632169}, 24 {\micron} \citep{babbedge20063701159, rujopakarn20107181171, rodighiero20105158} and total IR LFs \citep{lefloch2005632169, caputi200766097, magnelli200949657, rodighiero20105158}. These studies have found strong luminosity evolution with {\alphal} $\approx 3 - 5$ and moderate density evolution out to $z \sim 1$ implying IR galaxies were more luminous and numerous at higher redshifts than at $z = 0$. \cite{perezgonzalez200563082} analysed the IR galaxy LF to higher redshifts ($z \sim 3$) and found that the evolution remains constant from $z\sim1.2$ to 3.

\indent All these studies have shown an evolution of IR LFs with look-back time and the relative contribution from quiescent (L$_\mathrm{IR} < 10^{11}$ L$_\odot$), luminous IR galaxies (LIRGS: L$_\mathrm{IR} = 10^{11} - 10^{12}$ L$_\odot$) and ultraluminous IR galaxies (ULIRGS: L$_\mathrm{IR} = 10^{12} - 10^{13}$ L$_\odot$) to the cosmic SFR density.  At $z < 0.5$, the SFR is dominated by quiescent galaxies whereas at $z \sim 1$ LIRGs are responsible for $\sim 50$\% of the total IR density and dominate the star-forming activity beyond $z \geq 0.7$. \cite{perezgonzalez200563082} show that this evolution continues up to $z \sim 2.5$, and that ULIRGs play a rapidly increasing role for $z \geq 1.3$. \cite{caputi200766097} have shown that at $z \sim 2$, around 90\% of the IR luminosity density associated with star-formation is produced by LIRGs and ULIRGs. Recent results from analysis of IR LFs using data from the {\akari} satellite have found good agreement with the previous studies from {\iras}, {\iso} and \textit{Spitzer} (see \citealt{bethermin201051278, goto20105146}).

\indent Most of the previous LF work has been carried out in the mid-IR (MIR; $\lambda = 8 - 40 \mu$m), while studies of LFs at FIR wavelengths ($\lambda = 40 - 200 \mu$m) have been restricted to $z < 0.3$ (see \citealt{saunders1990242318, serjeant2004355813, takeuchi2006448525}). Studying the redshift evolution of LF at FIR wavelengths is vital because the CIB and the SEDs of most IR luminous galaxies both peak in this region of the IR spectrum. Furthermore, most of the progress in understanding the evolution of IR LFs has utilised \textit{Spitzer} 24 {\micron} observations, which are strongly dependent on the SED library as at high redshifts the 24 {\micron} channel sample shorter wavelengths. Further uncertainties in obtaining reliable estimates of the bolometric IR luminosities is introduced by redshifting of the polycyclic aromatic hydrocarbon (PAH) emission and silicate absorption features into the 24 {\micron} band at $z\gtrsim1$. 

\indent In this work, we investigate the evolution of the rest-frame 70{\micron} and total IR (TIR) LFs out to $z \sim 1.2$, and 160 {\micron} LF out to $z \sim 0.5$. We use optical and IR data from the XMM-LSS and Lockman Hole (LH) regions of the \textit{Spitzer} Wide-area InfraRed Extragalactic survey (SWIRE; \citealt{lonsdale2003115897, lonsdale200415454}). In \cite{babbedge20063701159} SWIRE optical and IR data were used to estimate photometric redshifts ({\photozs}) and construct LFs at 3.6, 4.5, 5.8, 8 and 24 {\micron} over the redshift range $0 < z < 2$ for galaxies and $0 < z < 4$ for optical quasi-stellar objects (QSOs). In the present work, the LFs are constructed using spectroscopic redshifts taken from the literature and our spectroscopic follow-up of SWIRE selected 70 {\micron} sources. Analysis using photo-$z$s work best at high redshifts, where fractional errors of $\Delta{z}/(1+z) \simeq 0.05$ would be acceptable. For example a source at $z = 1$, the uncertainty in $z$ would be a tolerable 10\%, whereas a source at $z = 0.3$ would have an uncertainty of $\sim 22$\%. 

\indent The paper is organised as follows: in Section 2, we present the optical and IR data and the spectroscopic redshifts from SWIRE-XMMLSS and SWIRE-LH regions used in this study. In Section 3 we describe the methodology used in calculating the LFs, starting with derivations of luminosities, parametrisation of the evolution of the LF with redshift and the estimation of SFR with the LF. The results of the LFs are presented in Section 4 including estimation of the evolution of the integrated SFR and further discussed in Section 5. We adopt a flat $\Lambda$CDM cosmology model with $H_{0}$ = 70 km s$^{-1}$ Mpc$^{-1}$ and $\Omega_{\Lambda}$ = 0.7.

\section{THE DATA}
\label{sect:thedata}
\subsection{Infrared Data}
\label{sect:infrareddata}
\indent The SWIRE survey \citep{lonsdale2003115897, lonsdale200415454} is one of the largest {\spitzer} legacy programmes covering a 49 deg$^2$ in six different fields (ELAIS-N1, ELAIS-N2, ELAIS-S1, CDFS, Lockman Hole and XMM-LSS) with both the IRAC (3.6 to 8 {\micron} channels) and MIPS (24 to 160 {\micron} channels) instruments. Typical 5$\sigma$ sensitivities are 3.7, 5.3, 48 and 37.7 $\mu$Jy in the IRAC 3.6, 4.5, 5.8 and 8{\micron} bands. For MIPS the 5$\sigma$ limits are  230 $\mu$Jy, 20 mJy, and 120 mJy at 24, 70 and 160 {\micron} \citep{surace2005371246}. The IRAC data were processed by the {\spitzer} Science Centre (SSC) IRAC pipeline and sources extracted using SExtractor \citep{bertin1996117393}. The MIPS 24 {\micron} data were processed by the SSC's MOPEX software and source extraction was performed using SExtractor. The MIPS 70 and 160 {\micron} data were also processed at SSC using the MOPEX software and source extraction carried out through the PRF fitting capabilities of MOPEX. Full details of the SWIRE data release can be found in \cite{surace2005371246}.

\indent The final data products consist of a cross-matched IRAC and MIPS 24 {\micron} catalogue and single-band catalogues at 24, 70 and 160 {\micron}. The IRAC and MIPS 24{\micron} catalogue consists of sources detected with a signal-to-noise ratio (S/N) $>$ 5 in one or more IRAC band and their 24 {\micron} associations with a S/N $>$ 3. In more than 90\% of the cases, the sources in the merged catalogue are within 1.5$''$ for IRAC channel-pairs and within 3$''$ for MIPS-24 to IRAC pairs and hence can be considered reliable. \cite{surace2005371246} evaluated the completeness of the IRAC data by comparing to deeper data in the ELAIS-N1 region taken as part of the Extragalactic First Look Survey (FLS) programme. The 95\% completeness level was calculated to be at 14, 15, 42 and 56 $\mu$Jy in the IRAC 3.6, 4.5, 5.8 and 8 {\micron} bands. The 24 {\micron} data is $\sim$ 97\% complete at 500 $\mu$Jy \citep{babbedge20063701159}. 

\indent The MIPS 70 and 160{\micron} sources single-band catalogues were matched to the IRAC and MIPS 24{\micron} bandmerged catalogues to produce a SWIRE IRAC and MIPS 7-band catalogue assuming a search radius of 9.3$''$ and 19.5$''$ for 70 and 160 {\micron} sources respectively, which correspond to the telescope limited resolution of the MIPS instrument (Vaccari et al., in prep). Thus, almost all of the 70 and 160 {\micron} sources are also detected at 24 {\micron}. The 70 and 160 {\micron} catalogues are 90\% complete down to 14 and 75 mJy for 70 and 160 {\micron} sources with S/N $>$ 3 (Vaccari et al., in prep). Full details on the data processing, completeness and reliability for the SWIRE 70 and 160 {\micron} observations will be provided in Vaccari et al. (in preparation). 

\begin{figure*}
\begin{center}
\vspace{0.5cm}
\hspace{0.3cm}
\subfigure{\label{fig:s70vsr}\includegraphics[height=5.5cm, width=8cm]{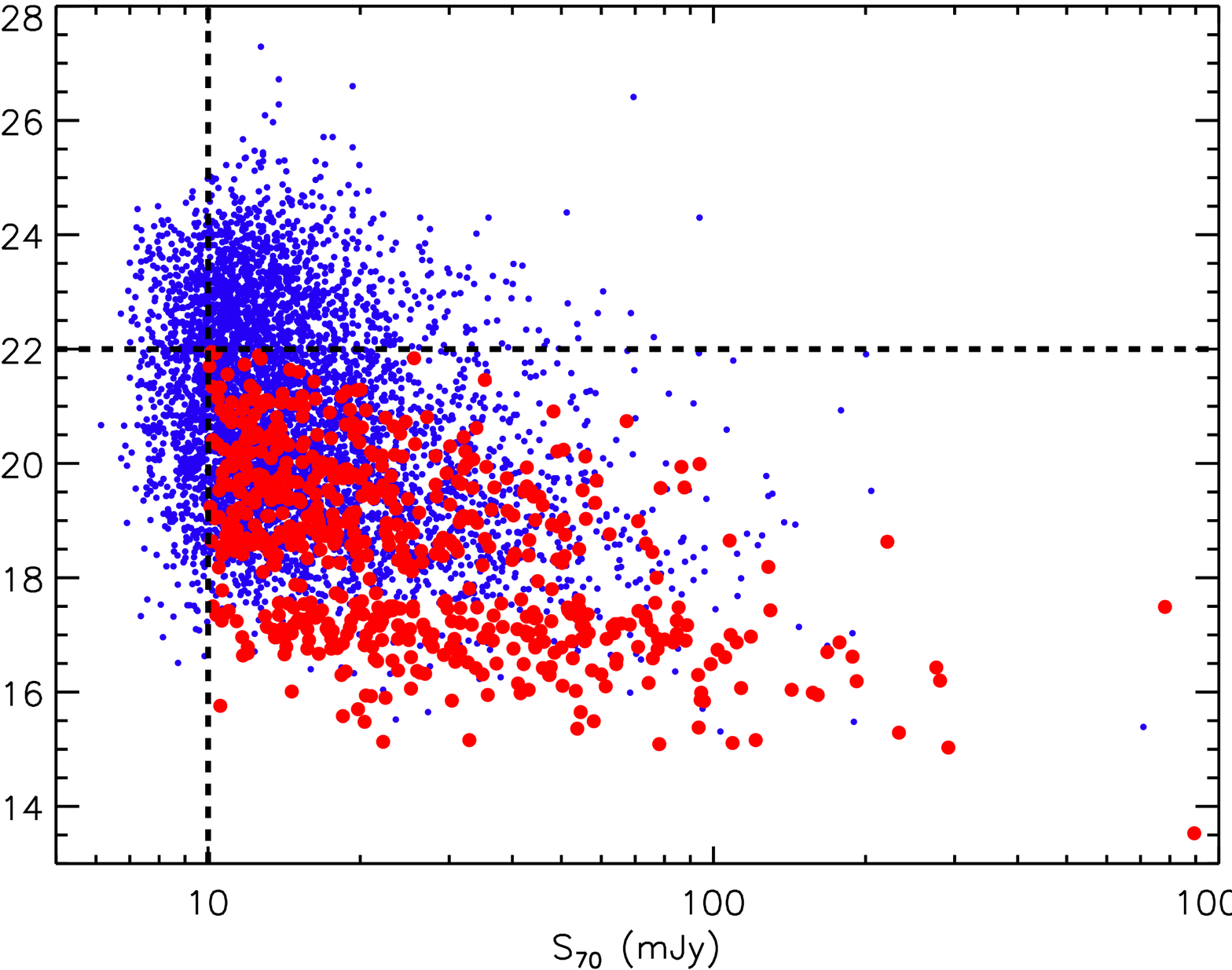}}
\hspace{0.5cm}
\subfigure{\label{fig:s160vsr}\includegraphics[height=5.5cm, width=8cm]{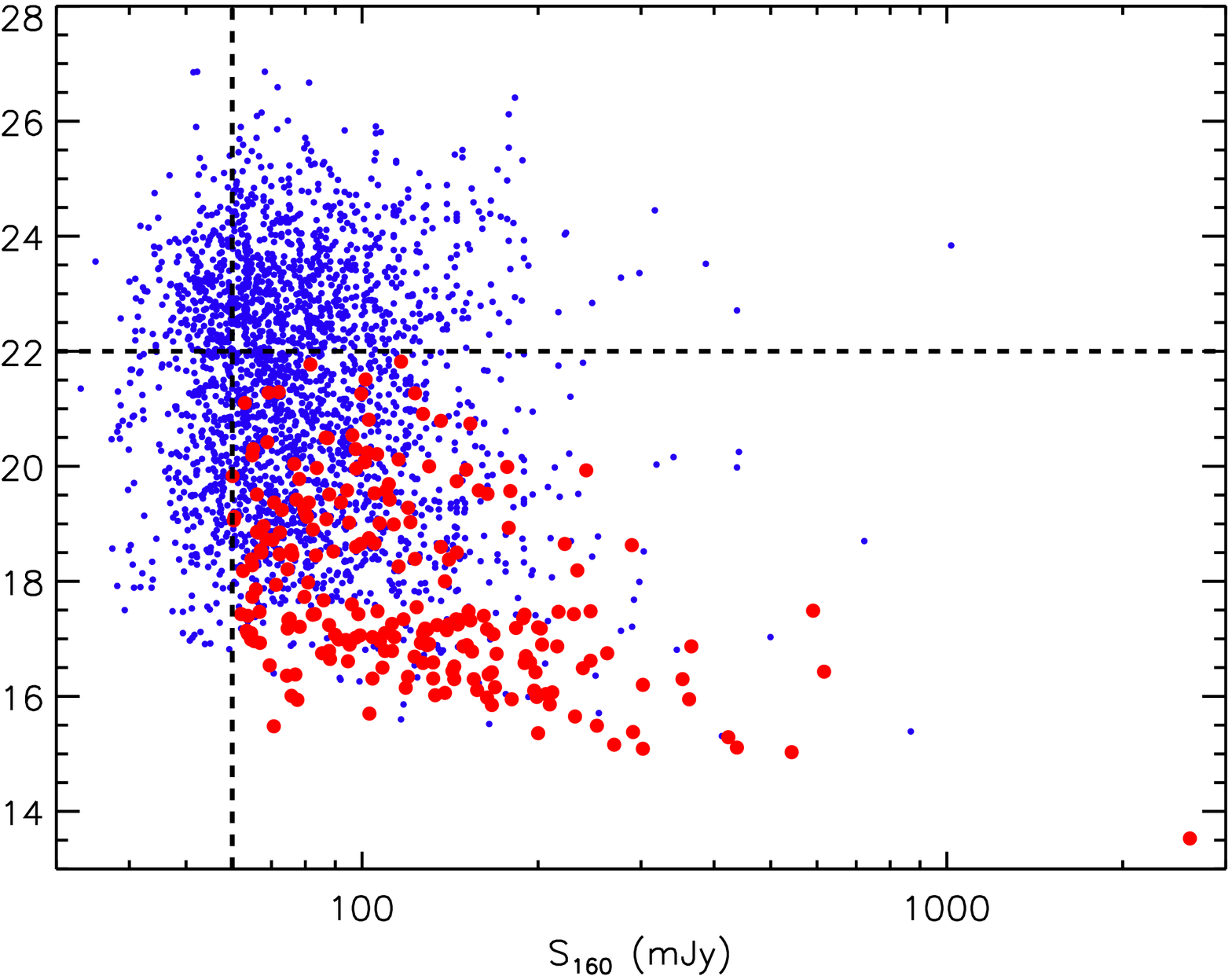}}
\end{center}
\caption{\textbf{Left}: r-band magnitude against 70 {\micron} flux. Blue dots are all 70 {\micron} detected sources with an optical counterpart in the LH and XMM-LSS. Filled red circles are the sources in the final 70 {\micron} sample. The vertical line marks the 70 {\micron} selection limit of 10 mJy and the horizontal line is at \textit{r} = 22. \textbf{Right}: r-band magnitude against 160 {\micron} flux. Blue dots are all 160 {\micron} detected sources with an optical counterpart in the LH and XMM-LSS . Filled red circles are the sources in the final 160 {\micron} sample. The vertical line marks the 160 {\micron} selection limit of 60 mJy and the horizontal line is at \textit{r} = 22.}
\label{fig:irfluxvsr}
\end{figure*}

\indent In this paper we have used data from the latest release of the SWIRE photometric redshift catalogue of \cite{rowanrobinson2008386687}. The catalogue contains photometric redshifts for over 1 million IR sources, estimated by combining the optical and IRAC 3.6 and 4.5 {\micron} photometry to fit the observed SEDs with a combination of galaxy and AGN templates \citep{babbedge20043701159, rowanrobinson20051291183}. Our analysis uses multiwavelength data from the XMM-LSS and Lockman Hole (LH) regions of the SWIRE survey. 

\subsection{Optical Data}
\label{sect:opticaldata}

Optical photometry is available for $>$ 70\% of the SWIRE area in at least three of the \textit{U}, \textit{g}, \textit{r}, \textit{i} and \textit{Z} photometric bands \citep{rowanrobinson2008386687, trichas2009399663}. {\spitzer}-optical cross-identifications (XID) was carried out  between the optical and the IRAC-24{\micron} catalogues using a search radius of 1.5$''$ \citep{rowanrobinson20051291183, surace2005371246}. The cross-identification process ensured that each SWIRE source only had one optical match. Completeness and reliability of the XID was investigated by \cite{surace2005371246}, which showed that essentially the {\spitzer}-optical XIDs are 100\% complete. The requirement that the sources be detected at both 3.6 and 4.5 {\micron} at S/N $\geq$ 5 appears to eliminate spurious sources effectively and give a high-reliability catalogueue \citep{rowanrobinson20051291183, surace2005371246}.

\indent The LH region centred on $\alpha$=
$10^\mathrm{h}45^\mathrm{m}$, $\delta$ = $+57^\mathrm{d}59^\mathrm{m}$ covers $\sim$ 10.5 deg$^{2}$. Optical photometry is available covering 7.53 deg$^2$ in the \textit{g, r} and \textit{i} bands were obtained using the MOSAIC camera on the 4m-Mayall Telescope at Kitt Peak National Observatory (KPNO). The 5$\sigma$ limiting magnitudes (Vega) are 25.1, 24.4, and 23.7 in the three bands for point-like sources \citep{berta2007467565}. \textit{U}-band photometry is available in a smaller 1.24 deg$^2$ region to 5$\sigma$ limiting magnitude (Vega) of 24.1. The data reduction was performed with the Cambridge Astronomical Survey Unit (CASU) pipeline \citep{irwin200145105}.


\indent The XMM-LSS field centred on $\alpha$ = $02^\mathrm{h}21^\mathrm{m}$, $\delta$ = $-04^\mathrm{d}30^\mathrm{m}$ and covers 9.1 deg$^\mathrm{2}$. Optical data is available for 6.97 deg$^2$ of XMM-LSS, which was observed as part of the Canada-France-Hawaii Telescope Legacy Survey (CFHTLS) in the \textit{u, g, r, i}, and \textit{z} bands to magnitude (Vega, 5$\sigma$ for a point like object) limits of: 24.9, 26.4, 25.5, 24.9 and 23.4 respectively\footnote{See http://www.cfht.hawaii.edu/Science/CFHTLS/}. The photometry was taken from \cite{pierre2007382279}. In addition there is 10-band photometry (\textit{ugrizUBVRI)} from the VIMOS VLT Deep Survey (VVDS) programme \citep{lefevre2004417839} covering 0.79 deg$^2$ and very deep 5-band photometry (\textit{BVRiz}) in 1.12 deg$^2$ of the Subaru XMM Deep Survey (SXDS; \citealt{furasawa20081761}).

\subsection{Sample Selection}
\label{sect:sampleselection}

\begin{table*}
\begin{center}
\caption[Summary of the sample selection used to study the 70 and 160 {\micron} LFs. ]{Summary of the sample selection used to study the 70 and 160 {\micron} LFs.}
\label{tab:samplesummary}
\begin{tabular}{cccccccccccc}
\hline\hline
Field 	& $\alpha$ (J2000) & $\delta$ (J2000) & Survey area & \multicolumn{4} {c} {70 {\micron} sources}  & \multicolumn{4} {c}{160 {\micron} sources}\\
	 	& & &  ({\sqrdeg}) & {$N_\text{det}$}$^a$ & {$N_\text{sel}$}$^b$ & {$N_\text{sp}$}$^c$ & {$N_\text{sp}$}$^d$ & {$N_\text{det}$}$^a$ & {$N_\text{sel}$}$^b$ & {$N_\text{sp}$}$^c$ & {$N_\text{sp}$}$^e$\\
	 	& & & 		    & 				         & 				    & 					  & \textit{z} $\leq1.2$  & 				    	& 				     & 				        & \textit{z} $\leq0.5$\\\hline
LH     	& $10^\text{h}45^\text{m}22^\text{s}$ & $+57^\text{d}59^\text{m}05^\text{s}$ & 7.53 & 4046 & 2159 & 354 & 343 & 1276 & 584 & 166 & 157\\
XMM-LSS & $02^\text{h}21^\text{m}20^\text{s}$ & $-04^\text{d}30^\text{m}00^\text{s}$ & 6.97 & 2358 & 1606 & 299 & 291 & 1604 & 739 & 81 & 64\\
\hline\hline
\end{tabular}\\
{$^a$Number of sources detected; $^b$Number of sources selected using the selection criteria; $^c$Number of sources with spectroscopic redshift; $^d$Number of sources with spectroscopic redshift below 1.2; $^e$Number of sources with spectroscopic redshift below 0.5}
\end{center}
\end{table*}

To study the 70 and 160 {\micron} LFs and their evolution, we select 1.) 70{\micron} sources with \textit{r} $<$ 22 mag and S$_{70} >10$ mJy; and 2.) 160 {\micron} sources with \textit{r} $<$ 22 mag and S$_{160} > $ 60 mJy. Furthermore, we require the sources to have measured spectroscopic redshifts. The adopted 70 and 160 {\micron} flux limits correspond to 70\% completeness of the SWIRE survey. The total IR LF is derived using the 70 {\micron} sample.
 
\indent The SWIRE photometric redshift catalogue of \cite{rowanrobinson2008386687} is the parent catalogue from which we select the 70 and 160 {\micron} sources. In LH there are 4046 and 1276 sources detected at 70 and 160 {\micron} respectively. Applying our selection criterion, 2159 70 {\micron} and 584 160{\micron} sources were selected. Spectroscopic redshifts are available for 354 70 {\micron} and 166 160 {\micron} sources, which were collated from literature. 

\indent In XMM-LSS, 2358 sources have 70 {\micron} detections. We selected 1606 70{\micron} sources, of which 299 have spectroscopic redshifts 
At 160 {\micron} we found 1604 sources with 826 brighter than the flux limits stated above. We have spectroscopic redshifts for 81 160 {\micron} sources. The spectroscopic redshifts in XMM-LSS were obtained from our spectroscopic follow-up programme conducted over 6 nights between November 2008-November 2009 at the WHT using the AF2/WYFFOS instrument. The details of the spectroscopic follow-up are presented in \cite{patel20114151738}. 

\indent We choose to construct the 70 {\micron} and the total IR LFs in the redshift range $0< z\leq1.2$ and the 160 {\micron} LF in the redshift range $0<z\leq0.5$. Thus, the final total sample of galaxies considered for the 70 {\micron} and total IR LFs consists of 634 (343 in LH and 291 in XMM-LSS) objects and for the 160 {\micron} LF consists of 221 (157 in LH and 64 in XMM-LSS) objects. We show in Figure \ref{fig:irfluxvsr} the \textit{r}-band magnitude as a function of the 70 {\micron} flux for all sources in the LH and XMM-LSS regions (filled blue circles) and the final spectroscopic sample (filled red circles). 

\begin{figure}
\begin{center}
\vspace{0.5cm}
\includegraphics[width=7.5cm, height=6cm]{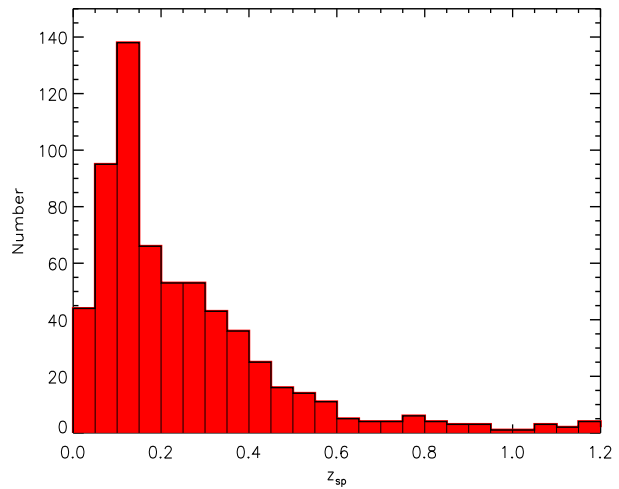}
\end{center}
\caption{Spectroscopic redshift distribution for the final 70 {\micron} sample. The 160 {\micron} sample is a subset of the 70 {\micron} sample restricted to $z<0.5$.}
\label{fig:speczdist}
\end{figure}

\indent A summary of the parent catalogue and the final sample is presented in Table.\ref{tab:samplesummary}. We display the spectroscopic redshift distribution of the final 70 {\micron} sample in Figure \ref{fig:speczdist}, while the final 160 {\micron} sample has the same redshift distribution for $z\leq0.5$, since all the 160 {\micron} detected sources are also detected at 70 {\micron}. The redshift distributions of the two fields is different because of large scale structures in the XMM-LSS field where the cluster distribution peaks around $z=0.3$ \citep{pacaud20073821289}, which also corresponds to the peak in the redshift distribution of our sample.

\section{INFRARED LUMINOSITIES}
\label{sect:seds}

\indent To determine the 70, 160 {\micron} and total IR LFs from our sample, we need to derive the rest-frame 70 and 160 {\micron} and the total IR luminosities. In order to do this, we model the SEDs for each source following the method described in \cite{rowanrobinson20051291183,rowanrobinson2008386687} used for the SWIRE photometric redshift catalogueue, using optical (at least 3 of the 5 optical \textit{U, g, r, i} and \textit{Z} bands) and IR photometry (\textit{Spitzer} IRAC 3.6 - 8 {\micron} and MIPS 24 - 160{\micron} bands). The SED fitting follows a two-stage approach, by first fitting the optical to near-IR (U to 4.5{\micron}) SED using the six galaxy and three AGN templates used by \cite{rowanrobinson2008386687}. 

\indent We calculate the IR excess by subtracting the galaxy model fit from the 4.5 to 24{\micron} data. We then fit the IR excess, 70 and 160{\micron} (for 83 70{\micron} sources) data points with the IR template of \cite{rowanrobinson20043511290, rowanrobinson20051291183, rowanrobinson2008386687}. The IR templates are derived from radiative transfer models dependent on interstellar dust grains, the geometry and the density distribution of dust. The IR templates are: 1.) IR 'cirrus': optically thin emission from interstellar dust illuminated by the interstellar radiation field; 2.) an M82 starburst; 3.) a more extreme Arp220-like starburst and 4.) an AGN dust torus. We also allow the sources to be fit by a mixture of: 1.) M82 starburst and cirrus, 2.) M82 starburst and AGN dust torus and 3.) Arp220 and AGN dust torus to properly represent the IR excess \citep{rowanrobinson1989238523, rowanrobinson20051291183}.

\begin{figure}
\begin{center}
\vspace{0.5cm}
\includegraphics[width=7.5cm, height=6cm]{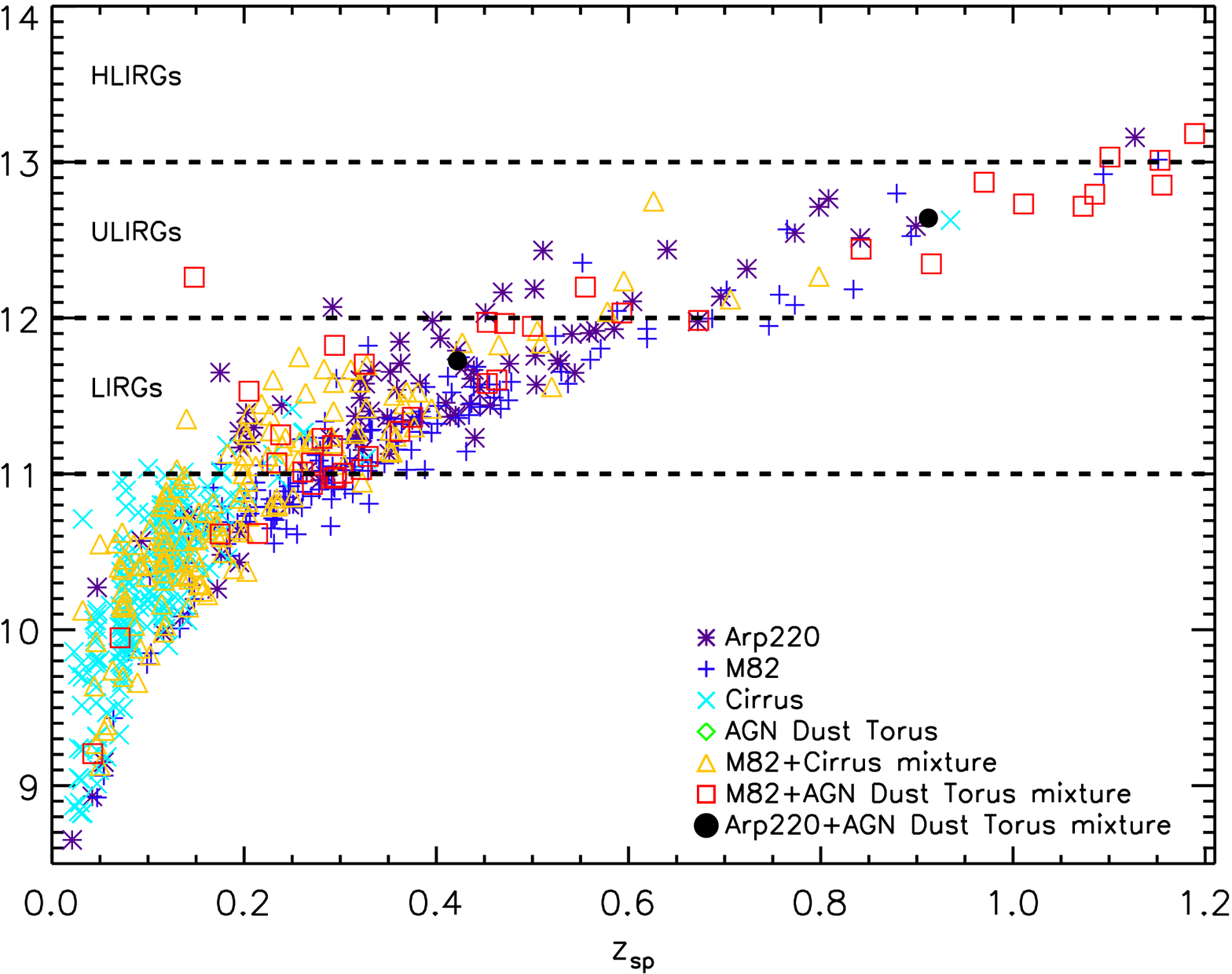}
\end{center}
\caption{Total IR luminosity ({\lir}) as a function of redshift for 70 {\micron} sources with flux above 9 mJy. The horizontal lines mark the divisions for LIRGs (10$^{11} \mathrm{L}_\odot < \mathrm{L_{IR}} < 10^{12} \mathrm{L}_\odot$), ULIRGs (10$^{12} \mathrm{L}_\odot < \mathrm{L_{IR}} < 10^{13} \mathrm{L}_\odot$) and HLIRGs (L$_\mathrm{IR} >10^{13} \mathrm{L}_\odot$).}
\label{fig:lirvsz}
\end{figure}

\begin{figure*}
\begin{center}
\vspace{0.7cm}
\hspace{0.2cm}
\subfigure{\label{}\includegraphics[height=6.cm, width=7.8cm]{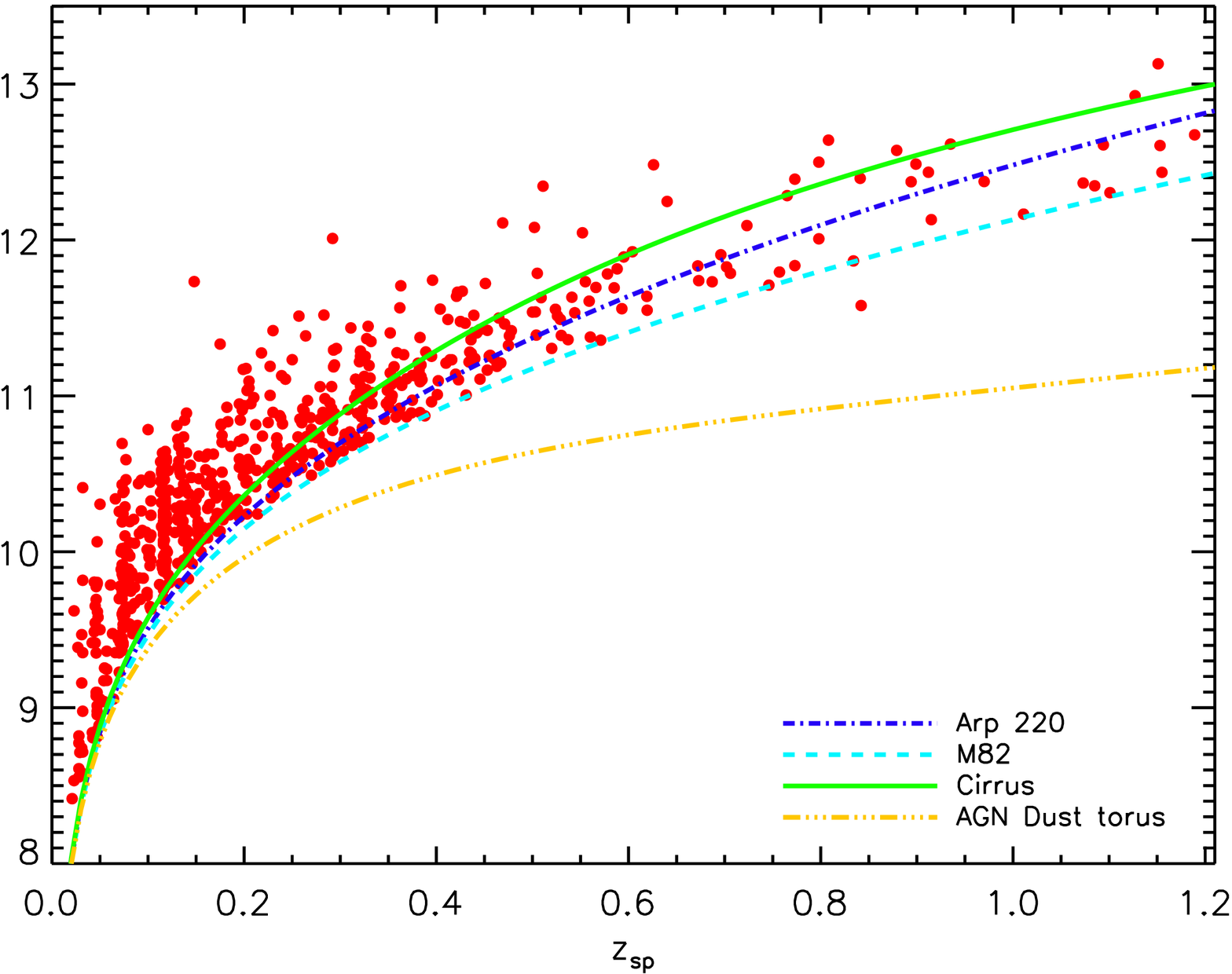}}
\hspace{0.8cm}
\subfigure{\label{}\includegraphics[height=6.cm, width=7.8cm]{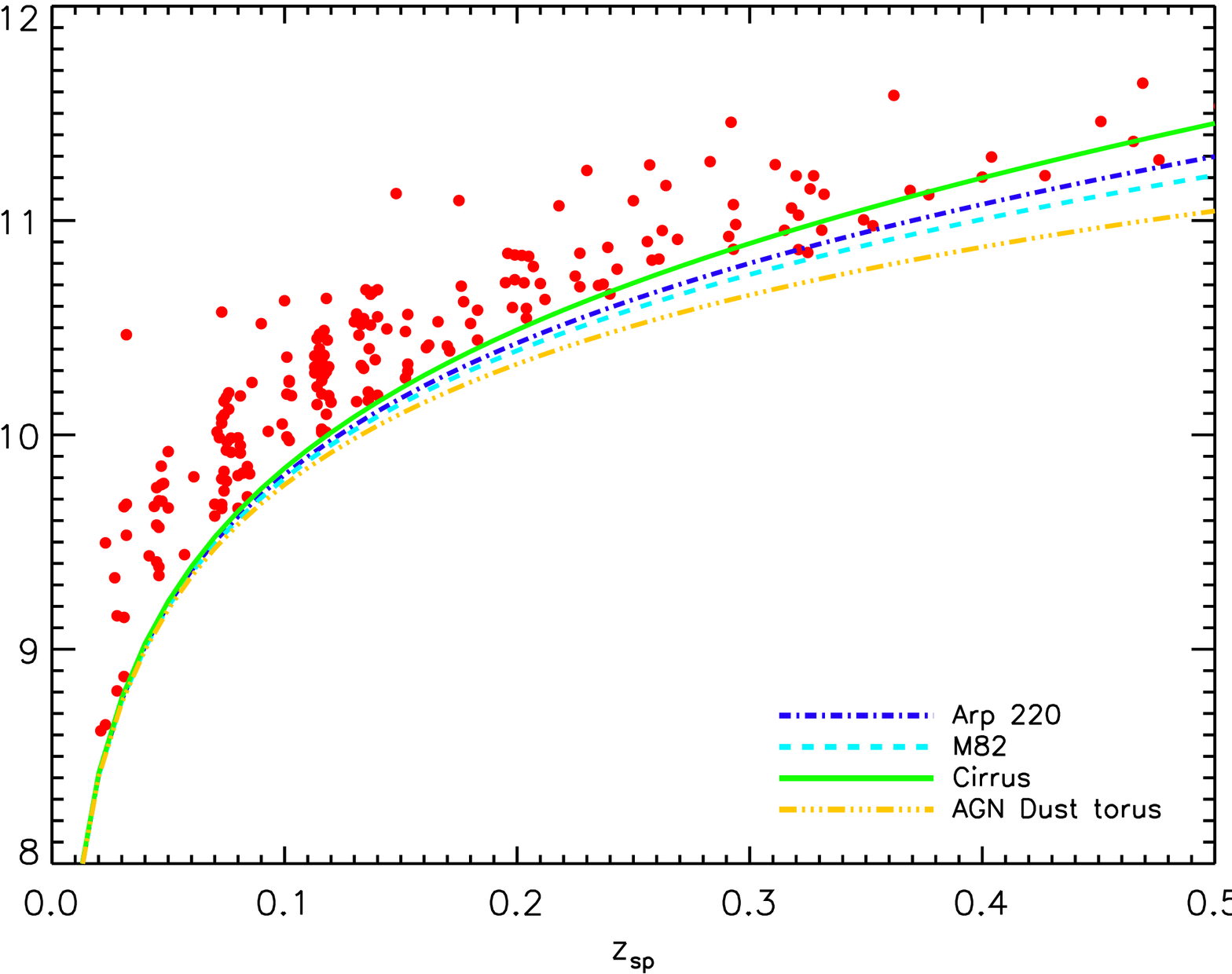}}
\end{center}
\caption{\textbf{Left}: Rest-frame 70 {\micron} luminosity ({\nulseventy}) as a function of redshift. The lines indicate as a function of redshift the 70 {\micron} rest-frame luminosity corresponding to an observed 70 {\micron} flux of 10 mJy with the 4 IR templates. \textbf{Right}:  Rest-frame 160 {\micron} luminosity ({\nulonesixty}) as a function of redshift. The lines indicate as a function of redshift the 160 {\micron} rest-frame luminosity corresponding to an observed 160 {\micron} flux of 60 mJy with the 4 IR templates.}
\label{fig:l70vsz}
\end{figure*}

\indent From the SED fitting we find all but 12 of the $z<1.2$ sample are fit with a galaxy template in the optical, which implies that the contamination by AGNs is low in our sample. Thus, we choose to include all sources in our analysis and do not separate the galaxies from AGNs. We illustrate in Figure \ref{fig:lirvsz} the distribution of total IR luminosity (estimated by integrating the SED models between 8 - 1000 {\micron}) as a function of redshift for the z $<$ 1.2 sample, with symbols representing the best fitting IR SED. The figure shows that in the IR the entire sample is dominated either by the starburst (Arp220 or M82-like) or the cirrus component, which is consistent with previous studies \citep{trichas2009399663, symeonidis20104031474, patel20114151738}. Figure \ref{fig:lirvsz} also shows that most of the 70 {\micron} sample are LIRGs and none of our sources have a pure AGN dust torus dominated IR SED. 

\indent We show in Figure \ref{fig:l70vsz} the rest-frame 70 {\micron} luminosity ({\nulseventy}) distribution as a function of redshift (left panel) and the rest-frame 160 {\micron} luminosity ({\nulonesixty}) distribution as a function of redshift (right panel). The 70 and 160 {\micron} luminosities were computed using K-corrections derived from the best fitting model templates. The uncertainty in the 70 and 160 {\micron} and {\lir} are typically 0.1 dex as stated in \cite{patel20114151738}. The SED model templates are also used to determine {\vmax}, which is the volume corresponding to the maximum redshift at which a source could be detected by the survey given the optical and IR limits set in Section \ref{sect:sampleselection}.

\section{METHODOLOGY}
\label{sect:irlfmethods}

In this section, we describe the methods used to determine the redshift evolution of the rest-frame 70, 160 {\micron} and TIR LFs, derived using the spectroscopic sample described in Section \ref{sect:sampleselection} and using the optical and IR SED model of each source described in the previous section. The LF, denoted as $\Phi(L)~\text{d}\log{L}$, defines the number of objects per comoving volume within a luminosity range ${\log}L, {\log}L+\text{d}\log{L}$. We construct the 70 {\micron} and total IR LFs in the redshift bins [0--0.2], [0.2--0.4], [0.4--0.8] and [0.8--1.2] and the 160 {\micron} LF in the redshift bins [0--0.2] and [0.2--0.5] using the {\vmax} method \citep{schmidt1968ApJ151393, avni1980235694}. We also develop a Bayesian method to study the evolution of the 70 and 160 {\micron} and the TIR LFs. In our LF estimation methods we accurately model the different completeness and selection effects affecting our data. The selection and incompleteness functions are described in Section \ref{sect:selfuncs} and the binned and parametric LF methods are discussed in Section \ref{sect:vmaxmethod} and Section \ref{sect:likelihoodfunction} respectively.

\subsection{Selection and Incompleteness Function}
\label{sect:selfuncs}

The data set we have used to estimate the FIR LFs was selected by imposing multivariate flux limits to determine whether a source is included in our sample. In addition we require each of the sources to then have a spectroscopic redshift. Here we use the 70 {\micron} LF selection functions as an example but the general procedure is the same for the total IR and 160 {\micron} LFs.

\indent The 70 {\micron} LF is calculated by selecting sources in the SWIRE catalogue that have S$_{70} > $ 10 mJy, $r <$ 22 mag and measured spectroscopic redshift. Therefore our first selection function considers the probability that a source of a given 70 {\micron} flux (which is a function of luminosity, $L$, and redshift $z$) is detected by the survey. This is denoted as $p$(det$|L, z$), and has been determined by Vaccari et al. (in preparation) for each field in the SWIRE survey using Monte Carlo simulations. The second selection function considers the probability that a given 70 {\micron} source with luminosity, $L$ at redshift $z$ is associated with an optical counterpart at $r <$ 22 mag and is defined as $p(r< 22|L, z$). This was quantified by taking the SWIRE catalogue and constructing a source count distribution for all sources detected at 70 {\micron} and then constructing a similar source count distribution for all 70 {\micron} sources with $r <$ 22 mag and the ratio is used to estimate $p(r< 22|L, z$).

\indent Finally, we take into account the probability that a given source to be characterised with a spectroscopic redshift. The spectroscopic incompleteness is primarily determined by the $r$ band magnitude and therefore they are added as weights in the computation of the LF (see Section \ref{sect:vmaxmethod} and Section \ref{sect:likelihoodfunction}). We define the weights, $w$, as the inverse of the spectroscopic completeness. The spectroscopic completeness is determined by constructing the source distribution for all sources with a measured spectroscopic redshift as a function of $r$ magnitude and dividing this by the source count distribution for all 70 {\micron} sources with S$_{70}>10$ mJy and $r<22$ as a function of $r$ magnitude.

\index The selection functions were estimated for LH and XMM-LSS regions separately and used in constructing the 70, 160 {\micron} and total IR LFs. The selection function, defined as \mbox{$p$(selected$|L, z$}), are combinations of the two selection functions:

\begin{align}
p\mathrm{(selected}|L, z) = p\mathrm{(det}|L, z)p(r<22|L, z)\label{eq:selectionfunction}
\end{align}


\noindent The selection function and the spectroscopic incompleteness were used to modify the LF methods discussed in the following section. 

\subsection{Binned Estimates}
\label{sect:vmaxmethod}

We use the {\vmax} method as the binned estimate which has the advantage that it allows direct computation of the LF from the data, without any parametric dependence or model assumption. We divided the sample into redshift bins selected to ensure adequate numbers of galaxies in each bin.  For each redshift bin, the LF is given by:

\begin{align}
\Phi(L) \mathrm{d}\log{L} = \left(\displaystyle\sum_{i}^{}w_i\times\frac{1}{V_{\mathrm{max}, i}}\right),\label{eq:lfvmax}
\end{align}

\noindent where $V_{\mathrm{max},i}$ is the comoving volume out to which the \textit{i}th galaxy could be observed, $w_i$ is the inverse of the spectroscopic incompleteness of the \textit{i}th galaxy. The comoving volume, $V_{\mathrm{max},i}$ is: 

\begin{align}
V_{\mathrm{max}, i}=\int_{z_\mathrm{min}}^{z_{\mathrm{max}, i}}p\mathrm{(selected|}L_i, z) \frac{\mathrm{d}V}{\mathrm{d}z}\mathrm{d}z, \label{eq:lfvmaxcalc}
\end{align}

\noindent where d$V$/d$z$ is the differential comoving element per unit solid angle \citep{hogg19995116}. Here $z_{\mathrm{max},i}$ corresponds to the maximum redshift at which the source could be detected by the survey given the optical flux limit ($r<22$) or the IR flux limits (S$_{70} >$ 10 mJy or S$_{160} >$ 60 mJy) and $z_{\mathrm{min}}$ is the lower limit of the redshift bin. $z_{\mathrm{max},i}$ was determined by using the optical/NIR and IR SED model of each source. The volume element integral in the LF calculation was weighted by the selection function to correct for the selection biases that are inherent in the spectroscopic catalogueue. The IR SED models described in Section \ref{sect:seds} are used in computing the selection function. 

\indent The associated rms error is given by:

\begin{align}
\sigma_{\Phi(L)}= \sqrt{\left(\displaystyle\sum_{i}^{}w_i^2\times\frac{1}{V_{\mathrm{max}, i}^2}\right)},\label{eq:lfvmaxerr}
\end{align}

\indent In order to accurately determine the uncertainty in the LF error we use Monte Carlo bootstrapping analysis to randomly resample the final spectroscopic catalogueue to generate 1000 realisations; each of these is analysed as described above and the root mean square (RMS) of the results is quoted as the error.

\subsection{Parametric Bayesian Method}
\label{sect:likelihoodfunction}

We use a Bayesian approach to determine the parametric LF, which requires prior knowledge of the appropriate functional form of the LF. As is the case for maximum likelihood (ML) methods \citep{sandage1979232352, marshall198326935} the parametric LF has the advantage over the {\vmax} method in that it is insensitive to any local clustering effect whereas the {\vmax} LF assumes a uniform number density throughout the observed volume and therefore is vulnerable to density inhomogeneities present in the survey \citep{wang201040135}. The advantage of using a Bayesian method to estimate the LF parameters over ML methods is that ML methods do not provide an estimate of the LF normalisation, which is often chosen to make the expected number of sources detected in a survey equal to the actual number of sourced detected. In addition, the confidence intervals on the LF parameters are derived assuming they have a Gaussian distribution which is not necessarily a good approximation for small sample sizes \citep{kelly2008682874}.

\indent In order to carry out a Bayesian analysis we first need to define the likelihood function, $p(\{d\}|\{\theta\}$), which is the probability of observing the data, $\{d\}$, for a given LF model, that is described by some parameters $\{\theta\}$. To do this we first define the probability of finding a source of a specific luminosity $L_i$ in the range $\log{L}, {\log}{L} + \text{d}\log{L}$ at a redshift $z_i$ in the range $z, z+\mathrm{d}z$ as:

\begin{align}
p(L, z|\{\theta\}) = \frac{\Phi(L,z|\{\theta\})p(\text{selected}|L, z)}{\lambda}\frac{\mathrm{d}V}{\mathrm{d}z},\label{eq:problz}
\end{align}

\noindent where $\lambda$ is the expected number of sources and is determined by:

\begin{align}
\lambda=\displaystyle\sum_\mathrm{fields}\iint\Phi(L,z|\{\theta\})p\mathrm{(selected|}L, z)\mathrm{d}\log{L}\frac{\mathrm{d}V}{\mathrm{d}z}\mathrm{d}z\label{eq:lambdapoiss}
\end{align}

\noindent The sum is taken over the fields present in our survey and the integrals are taken over all possible values of redshifts and luminosities. 

\indent We now write the likelihood function as the probability of observing $N$ objects, each with $L_i$ and $z_i$, drawn from the model LF as:

\begin{align}
p(\{d\}|\{\theta\}) &= p(N, \{L_i, z_i\}|\{\theta\}) \nonumber\\
&= p(N|\{\theta\})p(\{L_i, z_i\}|\{\theta\})
\end{align}

\noindent where $p(N|\{\theta\})$ is the probability of observing $N$ objects given the model LF and $p(\{L_i, z_i\}|\{\theta\})$ is the likelihood of observing a set of $L_i$ and $z_i$ given the model LF. We assume that the number of sources detected follows a Poisson distribution, where the expected number of detectable sources, $\lambda$, is given by Equation \ref{eq:lambdapoiss}. Thus the likelihood function is written as:


\begin{align}
p(N, \{L_i, z_i\}|\{\theta\})&=\frac{\lambda^Ne^{-\lambda}}{N!}\displaystyle\prod_{i=1}^Np(L_i, z_i|\{\theta\})\\
&= \frac{\lambda^Ne^{-\lambda}}{N!}\;\;\times\nonumber\\
&\displaystyle\prod_{i=1}^N\frac{\Phi(L,z|\{\theta\})p\mathrm{(selected|}L, z)}{\lambda}\frac{\mathrm{d}V}{\mathrm{d}z} \label{eq:lhbeforemod}
\end{align}

\indent We further modify the likelihood function by including the spectroscopic incompleteness by introducing a weighting factor, $w/\langle{w}\rangle$, for each object \cite[see][]{zucca1994269953, ilbert2005439863, aird2008387883}. The weights, added as exponents of the individual source likelihoods $p(\{L_i, z_i\}|\{\theta\})$ artificially reduce the size of error estimates. Therefore they are balanced by the average weight ($\langle{w}\rangle$) which do not effect the best fitting parameters \citep{aird2008387883}. Furthermore, $N$, the total number of objects is now $\sum{w_i}$, which gives the effective number of sources corrected for the spectroscopic incompleteness. Therefore the likelihood function in Equation \ref{eq:lhbeforemod} is:

\begin{align}
p(N, \{L_i, z_i\}|\{\theta\})&\propto\lambda^{\sum{w_i}}e^{-\lambda}\;\:\times\nonumber\\
&\displaystyle\prod_{i=1}^N\left\{\frac{\Phi(L,z|\{\theta\})p\mathrm{(selected|}L, z)}{\lambda}\frac{\mathrm{d}V}{\mathrm{d}z}\right\}^{\frac{w_i}{\langle{w}\rangle}}\nonumber\\\label{eq:lhfunc}
\end{align}

\noindent Note that in the absence of the spectroscopic incompleteness weights, $w_i$, Equation \ref{eq:lhfunc} reduces to the form presented in \cite{marshall198326935}.

\indent We perform Bayesian inference by combining the LF with a prior probability distribution, $p(\{\theta\})$ to compute the posterior probability distribution, $p(\{\theta\}|\{d\})$, given by Bayes' theorem:

\begin{align}
p(\{\theta\}|\{d\}) = \frac{p(\{d\}|\{\theta\})p(\{\theta\})}{\int{p(\{d\}|\{\theta\})p(\{\theta\})}d\theta},\label{eq:bayestheorem}
\end{align}

\noindent where the denominator is the Bayesian evidence and is determined by integrating the likelihood over the prior parameter space. 

\indent For parameter inference, the Bayesian evidence serves to normalise the posterior distribution and is vital for Bayesian model comparison. Calculating the Bayesian evidence is computationally expensive since it involves integration over $n$-dimensions for an $n$ parameter LF model. Therefore we use standard Markov chain Monte Carlo (MCMC) methods to perform a random walk through the parameter space to obtain random samples from the posterior distribution. We employed the Metropolis-Hastings algorithm \citep{metropolis1953211087, hastings19705797}, in which a proposal distribution is used to guide the variation of the parameters. The algorithm uses a proposal distribution which depends on the current state to generate a new proposal sample. We accept a step if the probability of the model given the new parameter values is higher and also at random intervals when the probability is lower in order to allow the fit to proceed downhill to avoid local minima \citep{ptak2007667826, kelly2008682874}. 

\indent We assume a flat prior distribution for each parameter and ignoring the normalising factor, the Bayesian evidence, the posterior distribution for each parameter is then given by:

\begin{align}
p(\{\theta\}|\{d\}) \propto p(\{d\}|\{\theta\})\label{eq:bayesparamest}
\end{align}

\noindent We produce three chains for each analysis of at least $2\times10^6$ iterations and adjust the parameter step sizes to achieve an acceptance ratio in the range 0.3 - 0.5. Finally we calculated the convergence $R$ statistic from \cite{gelman2004}, which should be $\leq1.2$ if the chain has converged. For all the parameters the $R$ value was $<1.1$.


\subsubsection{Luminosity Function Models}
\label{sect:70mulfspeczlfmodels}

The first IR LF constructed from {\iras} observations showed an excess in the number of galaxies at the high luminosity end \citep{soifer1987320238} from the value expected from the Schechter function \citep{schechter1976203297}. \cite{soifer1987320238} fit a double power-law model to the {\iras} 60 {\micron} LF which is adopted in our work. We use a continuous double power-law model  given by:

\begin{align}
\Phi(L|\{\theta\})=\phi^\ast\left[\left(\frac{L}{L^\ast}\right)^{\alpha}+\left(\frac{L}{L^\ast}\right)^{\beta}\right]^{-1}\label{eq:brokenpowerlaw}
\end{align}

\noindent Alternatively \cite{saunders1990242318} fit the {\iras} 60 {\micron} LF with a combination of a power-law and log-normal LF model:

\begin{align}
\Phi(L|\{\theta\})=\phi^\ast\left(\frac{L}{L^\ast}\right)^{(1-\alpha)}\exp\left[-\frac{1}{2\sigma^2}\log^2\left(1+\frac{L}{L^\ast}\right)\right].\label{eq:saunders90lfmodel}
\end{align}

\indent In both models, $L^\ast$ is the characteristic luminosity, $\phi^\ast$ is the LF normalisation and $\alpha$ is the power-law of the faint-end of the LF. $\beta$ in Equation \ref{eq:brokenpowerlaw} is the power-law index of the bright-end of the LF and $\sigma$ in Equation \ref{eq:saunders90lfmodel} gives the range over which the LF drops off.

\indent We estimate the parameters of the double power-law model, defined model 1, and the power-law and log-normal model, defined model 2. Uniform prior probability for all parameters were assumed with limits as follows: $-1\leq\alpha\leq2$, $0\leq\sigma\leq1$, $1\leq\beta\leq6$, $8\leq\log{L^\ast}\leq12$ and $-1\leq\log{\phi^\ast}\leq-4$.

\subsubsection{Luminosity Function Evolution Model}
\label{sect:lfevolmodesl}

Several studies \cite[see][]{lefloch2005632169, perezgonzalez200563082, caputi200766097, magnelli200949657, rujopakarn20107181171} of the IR luminosity functions, mainly at 24 {\micron} have shown that out to $z\sim1.2$, the while the luminosity and the number density of the LF evolves, the shape remains the same, that is $\alpha$, $\beta$ (in Equation \ref{eq:brokenpowerlaw}) or $\sigma$ (in Equation \ref{eq:saunders90lfmodel}) do not change. We parameterize the evolution of the FIR LFs in luminosity only:

\begin{equation}
\Phi(L, z|\{\theta\}) = \Phi\left[\frac{L}{f(z)}|\{\theta\}\right],
\label{eq:lfevolution}
\end{equation}

\noindent where $f(z) = (1+z)^\text{\alphal}$. Therefore in addition to the parameters listed in the previous section, {\alphal}, the luminosity evolution power-law index was also constrained. Uniform prior probability distributions were assumed for {\alphal} with limits: $1\leq\text{\alphal}\leq7$. Thus, for each LF model (double power-law and power-law and log-normal) five parameters were estimated.

\section{RESULTS}
\label{sect:results}

In this section, we present our determinations of the FIR LF and compare them to previous studies. We construct the rest-frame 70 {\micron} and the TIR LFs in redshift range $0<z\leq1.2$ and the rest-frame 160 {\micron} LF in the redshift range $0<z\leq0.5$. Finally using the TIR LF we derive the co-moving IR luminosity density and the cosmic star formation rate density as a function of redshift in the range $0<z\leq1.2$.

\subsection{Evolution of the Rest-frame 70 \boldmath${\mu}$m Luminosity Function}
\label{sect:70mulf}

\begin{figure*}
\begin{center}
\vspace{0.5cm}
\includegraphics[scale=0.48]{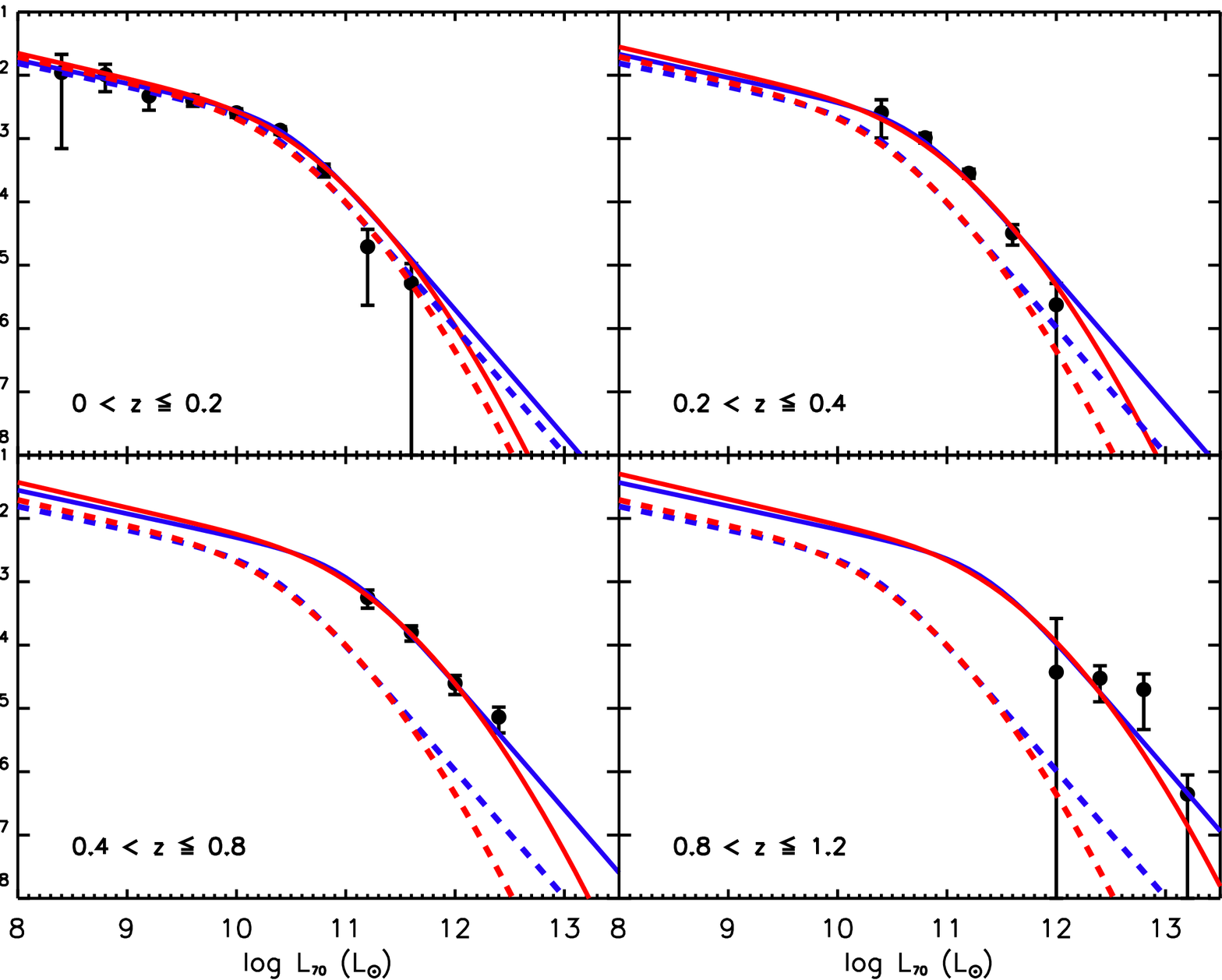}
\end{center}
\caption{70 {\micron} LF split into redshift bins $0 < z\leq0.2$, $0.2 < z\leq0.4$, $0.4 < z\leq0.8$ and $0.8 < z \leq 1.2$. The {\vmax} LF is shown as filled black circles, the solid blue line is the double power-law LF and the solid red line is the power-law and log-normal LF. The parametric LFs are displayed using the best-fit parameters listed in Table \ref{tab:lfbestfitparams} and evaluated at $z=0.1$, 0.3, 0.6 and 1. The dashed red and blue lines are the $z=0$ LFs.}
\label{fig:70mulfspecz70mulfs}
\end{figure*}

In Figure \ref{fig:70mulfspecz70mulfs}, we show the rest-frame 70 {\micron} LF constructed using the {\vmax} (black filled circles) and the parametric Bayesian (solid red and blue lines) methods described in Section \ref{sect:irlfmethods}. The parametric LFs are displayed using the best-fit (posterior mode) parameters given in Table \ref{tab:lfbestfitparams}. We also display the $z=0$ LF (dashed red and blue lines), which shows the rapid evolution of the 70 {\micron} LF, when compared with the binned and evolved parametric LFs. The parametric 70 {\micron} LF uses k-correction according to the M82 starburst SED.

\begin{table*}
\begin{center}
\begin{tabular}{ccccc}
\hline\hline
                                                                      &			 \multicolumn{4}{c}{$\log\Phi$ ({\mpcdex})}\\
{\loglseventy} ({\lsun})			     &           $0<{z}\leq0.2$             &      $0.2<{z}\leq0.4$              &       $0.4<{z}\leq0.8$            & $0.8<{z}\leq1.2$ \\
\hline
8.4                                                                 & $-1.96^{+0.29}_{-1.20}$ & 				        & 					&\\
8.8 							     & $-1.99^{+0.16}_{-0.27}$ & 				        & 					&\\
9.2  							     & $-2.33^{+0.15}_{-0.22}$ & 				        & 					&\\
9.6   							     & $-2.39^{+0.08}_{-0.10}$ & 				        & 					&\\
10.0							     & $-2.59^{+0.06}_{-0.07}$ &					        & 					&\\
10.4 							     & $-2.87^{+0.05}_{-0.06}$ & $-2.59^{+0.20}_{-0.40}$ & 				         &\\
10.8 							     & $-3.49^{+0.09}_{-0.11}$ & $-2.99^{+0.07}_{-0.08}$ & 					&\\
11.2 							     & $-4.71^{+0.27}_{-0.92}$ & $-3.55^{+0.07}_{-0.08}$ & $-3.25^{+0.12}_{-0.16}$ &\\
11.6 							     & $-5.28^{+0.31}_{-5.28}$ & $-4.49^{+0.13}_{-0.19}$ & $-3.80^{+0.10}_{-0.14}$ &\\
12.0 							     &                                             & $-5.62^{+0.34}_{-5.62}$ & $-4.60^{+0.13}_{-0.18}$ & $-4.43^{+0.85}_{-4.43}$\\
12.4 							     & 					      & 				       & $-5.14^{+0.16}_{-0.25}$ & $-4.52^{+0.20}_{-0.37}$\\
12.8 							     &					      & 				       & 				        & $-4.70^{+0.25}_{-0.63}$\\
13.2 							     & 					      & 				       & 				        & $-6.36^{+0.30}_{-6.36}$\\
\hline\hline
\end{tabular}
\end{center}
\caption{{\vmax} 70 {\micron} luminosity function values.}
\label{tab:70mulfvmaxvalues}
\end{table*}

\begin{table*}
\begin{center}
\begin{tabular}{lccccccc}
\hline\hline
		 			& \multicolumn{2}{c}{70 {\micron} LF} 				& \multicolumn{2}{c}{160 {\micron} LF} 					& \multicolumn{2}{c}{Total IR LF}\\
Parameter			& (a)						& (b)						& (a)						& (b)						& (a)						& (b)\\
\hline
{\loglstar} (L$_\odot$) 	& $10.35^{+0.12}_{-0.09}$ 	 & $9.53^{+0.10}_{-0.11}$		& $10.21^{+0.10}_{-0.08}$	& $9.85^{+0.18}_{-0.45}$		& $10.61^{+0.05}_{-0.13}$	& $9.71^{+0.19}_{-0.19}$\\
$\alpha$ 				& $0.37^{+0.10}_{-0.13}$ 		& $1.40^{+0.10}_{-0.09}$		& $0.40^{+0.17}_{-0.21}$		& $1.07^{+0.14}_{-0.58}$		& $0.34^{+0.10}_{-0.12}$ 		& $1.38^{+0.09}_{-0.12}$\\
$\beta$ 				& $2.00^{+0.13}_{-0.12}$	 	& --						& $3.79^{+0.36}_{-0.23}$		& --						& $2.09^{+0.14}_{-0.10}$		& --\\
$\sigma$				&  --					  	& $0.66^{+0.07}_{-0.03}$		& --	 					& $0.29^{+0.08}_{-0.05}$		& --				 		& $0.63^{+0.06}_{-0.04}$\\
{\logphistar} (Mpc$^{-3}$) & $-2.68^{+0.09}_{-0.14}$ 	& $-2.32^{+0.14}_{-0.06}$	& $-2.84^{+0.15}_{-0.11}$ 	& $-2.43^{+0.11}_{-0.16}$	& $-2.68^{+0.12}_{-0.10}$ 	& $-2.26^{+0.11}_{-0.12}$\\
{\alphal}				& $3.39^{+0.12}_{-0.22}$	 	& $3.41^{+0.18}_{-0.25}$		& $5.73^{+0.30}_{-0.62}$ 		& $5.53^{+0.28}_{-0.23}$		& $3.82^{+0.24}_{-0.20}$ 		& $3.82^{+0.25}_{-0.16}$\\
\hline\hline
\end{tabular}
\end{center}
\caption{Best-fitting parameters for of the 70 and 160 {\micron} and TIR LFs determined using the parametric Bayesian method. The errors include 68\% of the posterior probability. (a) double power-law LF model and (b) power-law and log-Gaussian LF model.}
\label{tab:lfbestfitparams}
\end{table*}

\indent Examination of the binned and parametric LFs in Figure \ref{fig:70mulfspecz70mulfs}, shows excellent agreement across all redshift bins. Comparison of the two LF models indicates that they are almost identical except at the brightest luminosities ({\lseventy} $>10^{12}$ {\lsun}), where the power-law and log-normal LF model has a rapid drop off when compared with the double power-law model. However, since there are no data points at these brightest luminosities, both LF models are a good description of the evolution of the 70 {\micron} LF. The shape of the faint-end of the LF is also consistent; $\alpha=0.37^{+0.10}_{-0.13}$ for model 1 and; $\alpha=1.40^{+0.10}_{-0.09}$ for model 2 as observed in the $0<z\leq0.2$ redshift bin. The luminosity evolution parameter, {\alphal}, which shows that the characteristic luminosity evolves rapidly as a function of redshift is almost identical for the two LF models ({\alphal} $=3.39^{+0.12}_{-0.22}$ and {\alphal} $=3.41^{+0.18}_{-0.25}$ for model 1 and 2 respectively). The best fit evolutionary parameter is consistent with studies performed at several IR wavelengths, in particular at 24 {\micron}, which find strong luminosity evolution with $\text{\alphal}=3-5$ \citep[see][]{lefloch2005632169, babbedge20063701159, magnelli200949657, rujopakarn20107181171}. We find that both parametric models can describe the evolution of the 70 {\micron} LF. 

\begin{figure}
\begin{center}
\vspace{0.5cm}
\hspace{0.4cm}
\includegraphics[scale=0.35]{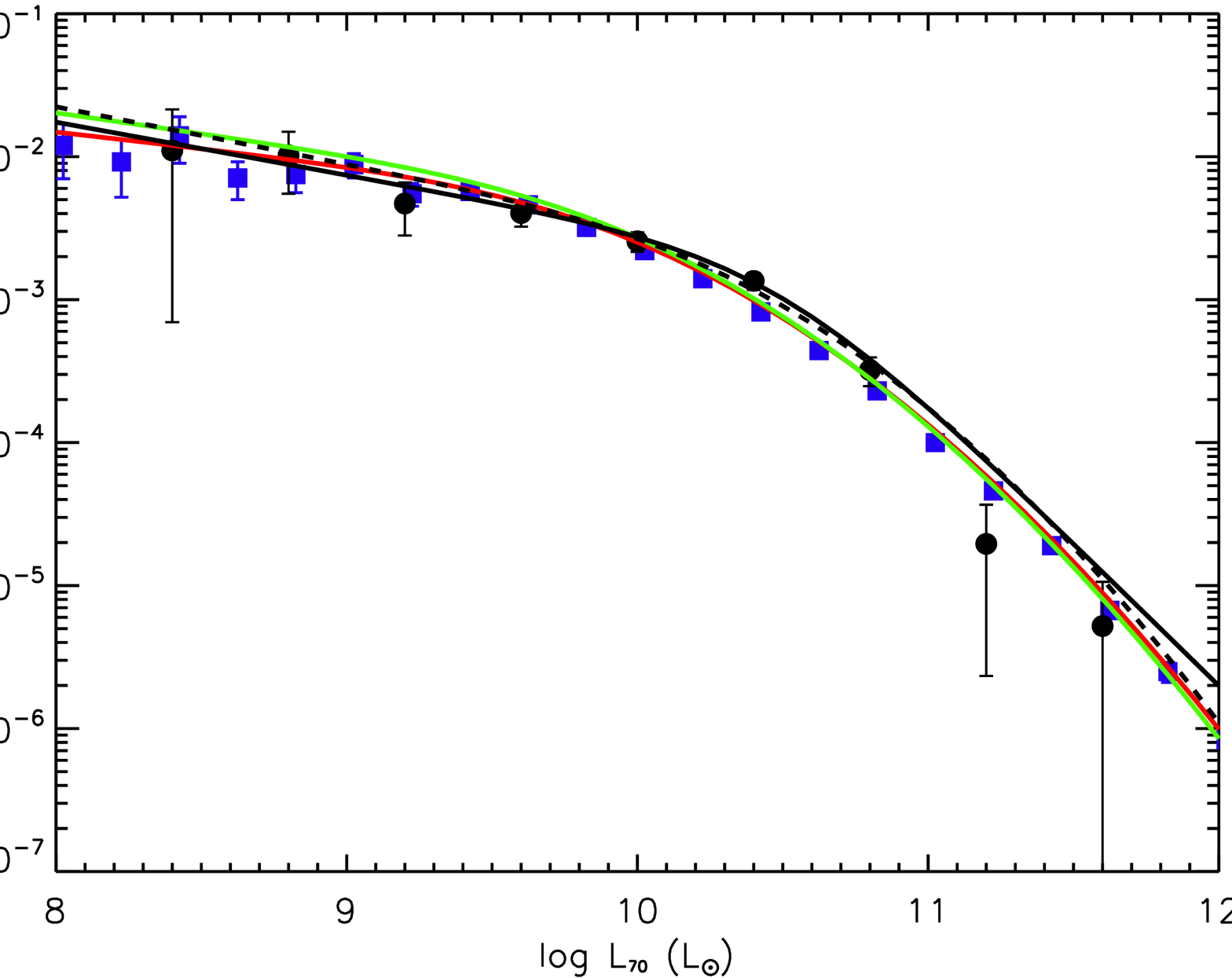}
\end{center}
\caption{The LLF ($0<z\leq0.2$) at 70 {\micron} determined using {\vmax} (filled black circles) and the parametric Bayesian method (solid and dashed black lines) evaluated at $z=0.1$. The solid black line is the double-power law model and the dashed black line is the power-law and log-normal model. Binned IRAS 60 {\micron} LF from \citet{saunders1990242318} is shown in filled blue squares and the parametric 60 {\micron} LF from \citet{takeuchi200358789} (solid red line) and \citet{wang201040135} (solid green line) are corrected for the bandpass differences assuming an M82 type starburst SED.}
\label{fig:70mullf}
\end{figure}

\indent We shown in Figure \ref{fig:70mullf}, the SWIRE 70 {\micron} local luminosity function (LLF) constructed in the redshift bin $0<z\leq0.2$, is compared to the bandpass corrected (assuming an M82 starburst SED) {\iras} 60 {\micron} LFs determined by \cite{saunders1990242318} (filled blue square), \cite{takeuchi200358789} (solid red line) and \cite{wang201040135} (solid green line). Comparison of the LFs shows that the overall shape of the parametric and binned 70 {\micron} LF is well matched to the bandpass corrected 60 {\micron} LFs. \cite{takeuchi200358789} {\iras} 60 {\micron} LF report, {\loglstar} ({\lsun}) $=9.07\pm0.09$, $\alpha=1.23\pm0.04$, $\sigma=0.72\pm0.01$ and {\logphistar} ({\mpcdex}) $=-2.05\pm0.05$ while \cite{wang201040135} determine, {\loglstar} ({\lsun}) $=9.10$, $\alpha=1.29$, $\sigma=0.72$ and {\logphistar} ({\mpcdex}) $=-2.05$. At the faint-end, the 70 {\micron} LF is steeper with $\alpha=1.40^{+0.10}_{-0.09}$ for model 2, while $\sigma = 0.66^{+0.07}_{-0.03}$, in excellent agreement with the value determined from {\iras} surveys. 

\indent We have used the k-correction given by the M82 starburst SED template in order to determine $p(\mbox{selected}|L, z)$ in the likelihood function because a large fraction of the 70 {\micron} population are fit with this template (see Section \ref{sect:seds} and Figure \ref{fig:lirvsz}). Although using a k-correction of a pure Arp220 starburst or a pure cirrus SED template results in different values for {\alphal}, we still find that the 70 {\micron} LF evolves rapidly in luminosity ({\alphal} $>$ 3). Studying the LF evolution of each SED component will be further investigated in future studies (see Section \ref{sect:swirelffutureprospects}). 

\subsection{Evolution of the Rest-frame 160 \boldmath${\mu}$m Luminosity Function}
\label{sect:160mulf}

\begin{figure*}
\begin{center}
\vspace{0.5cm}
\includegraphics[scale=0.48]{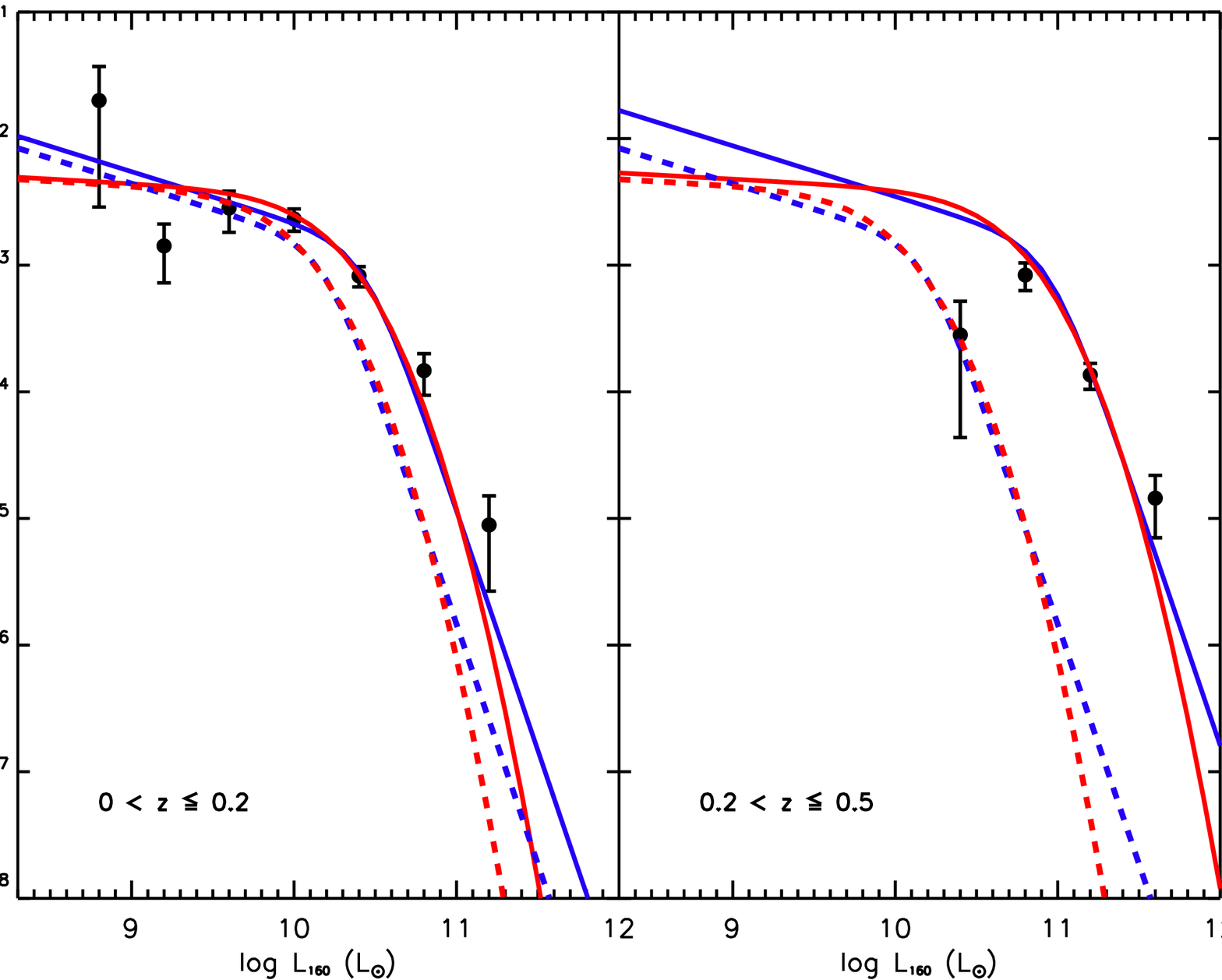}
\end{center}
\caption{160 {\micron} LF split into redshift bins $0 < z\leq0.2$ and $0.2 < z\leq0.5$. The {\vmax} LF is shown as filled black circles, the solid blue line is the double power-law LF and the solid red line is the power-law and log-normal LF. The parametric LFs are displayed using the best-fit parameters listed in Table \ref{tab:lfbestfitparams} and evaluated at $z=0.1$ and 0.5. The dashed red and blue lines are the $z=0$ LFs.}
\label{fig:160mulfevolution}
\end{figure*}

The rest-frame 160 {\micron} LF constructed using the {\vmax} (filled black circles) and the parametric Bayesian (solid red and blue lines) methods is shown in Figure \ref{fig:160mulfevolution}. The parametric LFs are displayed using the best-fit parameters given in Table \ref{tab:lfbestfitparams} and the {\vmax} values are listed in Table \ref{tab:70mulfspecz160mulfvmaxvalues}. The $z=0$ LF is also displayed (dashed red and blue lines), which show the rapid evolution of the 160 {\micron} LF, when compared with the binned and the evolved parametric LFs. The parametric Bayesian 160 {\micron} LF was calculated using a k-correction given by a mixture of M82 starburst (50\%) and cirrus template (50\%).

\begin{table*}
\begin{center}
\vspace{0.2cm}
\begin{tabular}{ccc}
\hline\hline
					& \multicolumn{2}{c}{$\log\Phi$ ({\mpcdex})}\\
{\loglonesixty} ({\lsun}) 	& 		$0<{z}\leq0.2$   	  & 			 $0.2<{z}\leq0.5$  \\
\hline
8.8 					& $-1.70^{+0.26}_{-0.84}$ &\\
9.2 					& $-2.85^{+0.17}_{-0.29}$ &\\
9.6 					& $-2.55^{+0.13}_{-0.19}$ &\\
10.0 					& $-2.63^{+0.08}_{-0.10}$ &\\
10.4 					& $-3.08^{+0.07}_{-0.09}$ & $-3.55^{+0.27}_{-0.81}$\\
10.8 					& $-3.83^{+0.13}_{-0.19}$ & $-3.08^{+0.10}_{-0.12}$\\
11.2 					& $-5.05^{+0.23}_{-0.52}$ & $-3.87^{+0.09}_{-0.11}$\\
11.6 					& 					 & $-4.84^{+0.18}_{-0.31}$\\
\hline\hline
\end{tabular}
\end{center}
\caption[The {\vmax} 160 $\mu$m LF values]{The {\vmax} 160 $\mu$m LF values.}
\label{tab:70mulfspecz160mulfvmaxvalues}
\end{table*}

\indent The binned and parametric LFs shown in Figure \ref{fig:160mulfevolution} are consistent with each other except at {\lonesixty} $>10^{11}$ {\lsun} where both the parametric LFs show a faster drop-off. In fact the double power law LF model shows a much better agreement with the {\vmax} LF in both redshift bins at {\lonesixty} $>10^{11}$ {\lsun} implying that the double power law LF model may be a better description of FIR LFs. The 160 {\micron} LF evolves in luminosity with {\alphal} $=5.73^{+0.29}_{-0.62}$ for model 1 and {\alphal} $=5.53^{+0.28}_{-0.23}$ for model 2, which is stronger than the value found for the evolution of the 70 {\micron} LF or from {\spitzer} 24 {\micron} studies \citep[see][]{lefloch2005632169, babbedge20063701159, magnelli200949657, rujopakarn20107181171}. \cite{takeuchi2006448525} however find that the \textit{ISO} 170 {\micron} LF evolves with {\alphal} $= 5.0^{+2.3}_{-0.5}$, being entirely consistent with the evolution of 160 {\micron} LF. Recently, \cite{dye201051810} used data from the H-ATLAS survey to find that the rest-frame 250 {\micron} luminosity density evolves at a rate proportional to $(1+z)^{7.1^{+2.1}_{-1.4}}$ to $z\simeq0.2$ in agreement with the evolution of 160 and 170 {\micron} LFs. 

\indent Our results show that the 160 {\micron} LF evolves more rapidly than the 70 {\micron} LF suggesting that `cooler' galaxies evolve more rapidly than `warmer' galaxies (the 160 {\micron} LF was determined using the k-correction given by a mixture of M82 starburst and cirrus template whereas the 70 {\micron} LF was determined using the k-correction given by an M82 starburst template). This is in keeping with the work of \cite{symeonidis2011411983} who find that `cold' galaxies evolve more rapidly than `warmer' galaxies over a period of $0.1<z<1$. \cite{dunne20114171510} have shown that the evolution of the 250 {\micron} LF out to $z=0.5$ is driven in part by evolution in the dust mass and an increase in the luminosity or space density of cooler galaxies. They conclude that the evolution of the dust mass points to an enhanced supply of gas for star formation at earlier cosmic epochs.

\begin{figure}
\begin{center}
\vspace{0.4cm}
\includegraphics[height=6.5cm, width=7cm]{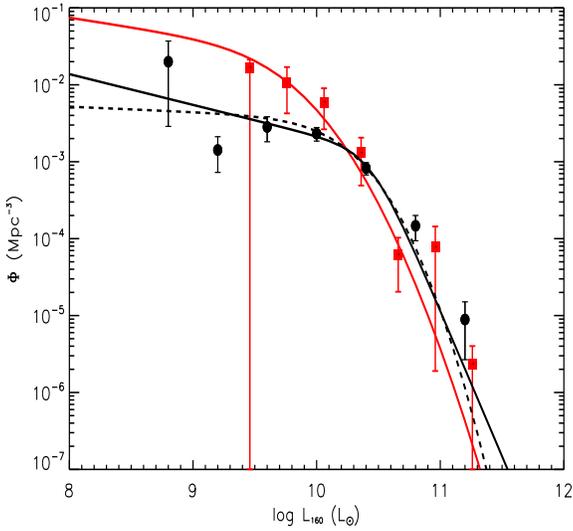}
\end{center}
\caption{The LLF ($0<z\leq0.2$) 160 {\micron} LF determined using {\vmax} (filled black circles) and the parametric Bayesian method (solid and dashed black lines) evaluated at $z=0.1$. The solid black line represents the double-power law model and the dashed black line is the power-law and Gaussian model.The parametric (red solid line) and non-parametric (red filled squares) \textit{ISO} 170 {\micron} LFs from \citet{takeuchi2006448525} are shown for comparison.}
\label{fig:160mullf}
\end{figure}

\indent We compare in Figure \ref{fig:160mullf} our 160 {\micron} LLF with the \textit{ISO} 170 {\micron} LF from \cite{takeuchi2006448525} constructed using 55 galaxies at $z < 0.3$. The 170 {\micron} binned LF shows a reasonable agreement within the error bars with the 160 {\micron} LF except at {\lonesixty} $\lesssim 10^{10}$ {\lsun} where there is a significant discrepancy between the two LFs, which is most likely due to the small number of objects in the 170 {\micron} sample. \cite{takeuchi2006448525} use the power-law and log-normal LF model and conclude that the parametric form (solid red line in Figure \ref{fig:160mullf}) underestimates the bright end of the LF, similar to the findings reported here. The combination of these results indicates that the double power-law model is the favoured analytic form for FIR LFs.

\subsection{Total Infrared Luminosity Function}
\label{sect:totalirlf}

In this section, we compute the total IR luminosity function of galaxies using the observed 70 {\micron} data. The total IR luminosity of a star forming galaxy provides a direct estimate of current star formation activity because the IR emission is the reprocessed UV/optical radiation produced by young stars. L$_\mathrm{IR}$ can be converted to SFR using the relationship provided by \cite{kennicutt199836189}:
\begin{equation}
\mathrm{SFR}\: (\mathrm{M_\odot\:yr^{-1}}) = 1.72\times10^{-10}~\mathrm{L_{IR}}\: \mathrm{L_\odot},
\label{eq:sfrk98}
\end{equation}

\noindent where L$_\mathrm{IR}$ is estimated by integrating for each source the best fit SED in the interval $8-1000 \mu$m (see Section \ref{sect:seds}). We can then use Equation \ref{eq:sfrk98} and the total IR LF to derive an estimate of the IR comoving energy density and the cosmic star formation rate density (CSFRD) up to z $\sim$ 1.2 and compare these results with other CSFRD calibrators.

\begin{figure}
\begin{center}
\vspace{0.5cm}
\includegraphics[height=5.5cm, width=7cm]{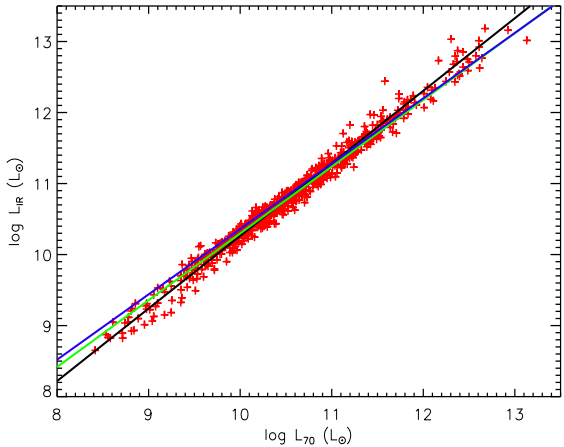}
\end{center}
\caption{Correlation between rest frame 70 {\micron} luminosity ($\log$ {\nulseventy}) and total IR luminosity ({\loglir}). The black solid line is our linear relationship determined using a Bayesian linear regression method. We also show for comparison the linear relation from \citet{symeonidis20083851015} (blue solid line) and \citet{bavouzet200847983} (green solid line).}
\label{fig:nul70vlir}
\end{figure}

\indent Previous studies of the the total IR LFs have largely been conducted at MIR wavelengths, with the greatest progress made at the {\spitzer} 24 {\micron} band where MIPS is most sensitive \citep[see.][]{lefloch2005632169, perezgonzalez200563082, babbedge20063701159, caputi200766097, bethermin201051643, rodighiero20105158}. The work presented here follows a similar approach to others in the literature, which relies on the conversion of $\mbox{L}_\nu$ to L$_\mathrm{IR}$, usually calibrated using SED templates of local IR galaxies \citep{chary2001556562} or semi-empirical SEDs \citep{dale2002576159, lagache2003338555}. Several authors \citep{elbaz2002384848, appleton2004154147} and in paritucular \cite{bavouzet200847983}, have shown that the extrapolation of the local $\mbox{L}_\nu-\mathrm{L_{IR}}$ remains reliable up to $z=1.1$ for LIRGs and $z\sim$ 2 for ULIRGs. Furthermore, the wavelength closest to the peak of far-IR emission provides the most accurate estimator of L$_\mathrm{IR}$ \citep{bavouzet200847983}. 

\indent We use the SED templates of \cite{rowanrobinson20043511290, rowanrobinson20051291183, rowanrobinson2008386687} and therefore to check the consistency of our $\mbox{L}_{70}-\mathrm{L_{IR}}$ correlation, we compared our results with those of \cite{bavouzet200847983} and \cite{symeonidis20083851015}. In Figure \ref{fig:nul70vlir} we show the relationship between $\log$ {\lseventy} and {\loglir} for all sources at $z<1.2$ in our sample (black solid line) with the relations of \cite{bavouzet200847983} (green solid line) and \cite{symeonidis20083851015} (blue solid line). The three correlations show excellent agreement with a mean scatter of $\sim 0.1$ dex between our correlation and the other two in the luminosity range $9.5\lesssim\log\text{L}_{70}\lesssim12.5$. Therefore we choose to use our $\mbox{L}_{70}-\mathrm{L_{IR}}$ correlation in constructing the total IR LF.

\indent To calculate the {\vmax} LF, we follow the same method as described in Section \ref{sect:vmaxmethod} and use the 70 {\micron} selection function for each source. For the Bayesian analysis we use our linear relationship between $\mbox{L}_{70}-\mathrm{L_{IR}}$ to convert L$_\mathrm{IR}$ to {\lseventy} to determine the selection function $p(\mathrm{selected}|L, z)$ in the likelihood function. Thus, the Bayesian analysis is dependent on the values of the linear relationship shown in Figure \ref{fig:nul70vlir}. We perform a test by estimating the parameters of the local total IR LF using the correlations of \cite{bavouzet200847983} and \cite{symeonidis20083851015} and found all parameters to be almost identical. 

\subsubsection{Evolution of the Total Infrared Luminosity Function}
\label{sect:evolutionoftirlf}

\begin{figure*}
\begin{center}
\vspace{0.7cm}
\includegraphics[scale=0.55]{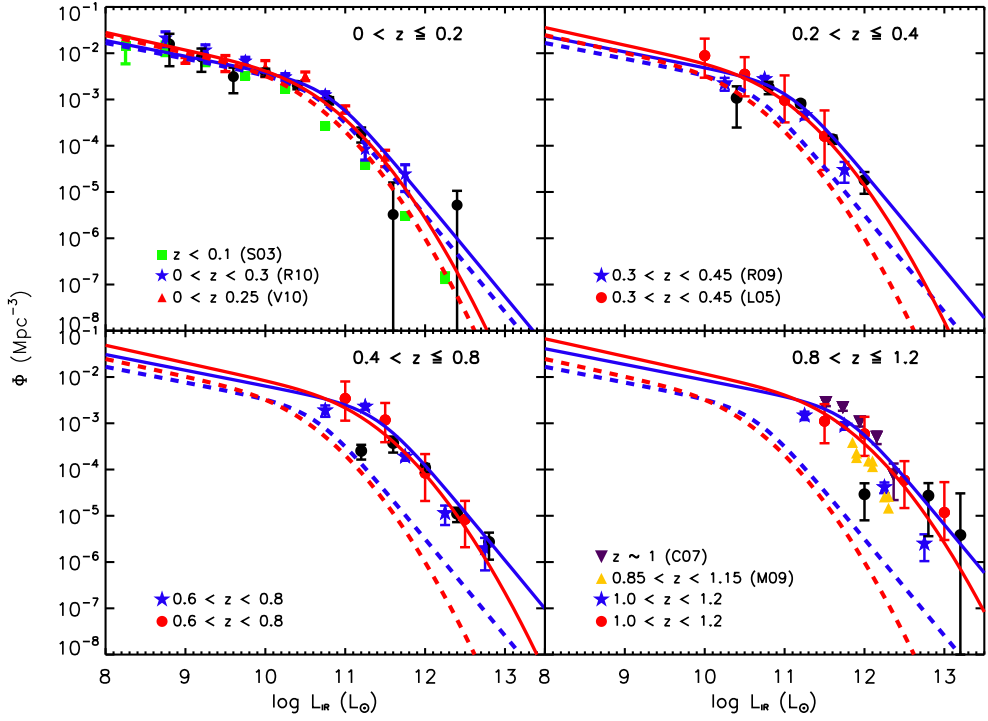}
\end{center}
\caption{TIR LF split into redshift bins $0 < z \leq 0.2$, $0.2 < z \leq 0.4$, $0.4 < z  \leq 0.8$ and $0.8 < z \leq 1.2$. Filled black circles are the {\vmax} estimates and the black solid line is the parametric Bayesian LF displayed using the best-fit parameters given in Table \ref{tab:lfbestfitparams}  at $z=0.1$, 0.3, 0.6 and 1. Filled red filled circles are from \citet{lefloch2005632169}, filled purple upside down triangles are from \citet{caputi200766097}, filled orange filled upside down triangles are from \citet{magnelli200949657} and filled blue filled stars are from \citet{rodighiero20105158}. The $z=0$ LF is shown in each panel (black dashed line).}
\label{fig:tirlfevolution}
\end{figure*}

In Figure \ref{fig:tirlfevolution}, we display the TIR LF constructed using the {\vmax} (filled black circles) and the parametric Bayesian (solid black line) methods. The parametric LFs are displayed using the best-fit parameters given in Table \ref{tab:lfbestfitparams} and the {\vmax} values are listed in Table \ref{tab:70mulfspecztirlfvmaxvalues}. The $z=0$ LF is also displayed (dashed black line), which show the rapid evolution of the TIR LF, when compared with the binned and the evolved parametric LFs. The parametric TIR LF was calculated using the k-correction given by the M82 starburst SED.

\indent Comparison of the {\vmax} and parametric LFs in Figure \ref{fig:tirlfevolution} shows good agreement across all redshift bins for both LF models.  We find the evolution of the TIR LF is independent of the two LF models used ({\alphal} $=3.82^{+0.24}_{-0.20}$ for model 1 and {\alphal} $=3.82^{+0.25}_{-0.16}$ for model 2), similar to the results of the 70 and 160 {\micron} LFs. Thus the choice of the LF model does not affect the conclusion that the TIR LF which shows strong evolution in luminosity. The global evolution parameter of the TIR LF is consistent with values determined by previous studies, which typically find {\alphal} $\sim 3.5\pm0.5$ \citep{caputi200766097, lefloch2005632169, magnelli200949657} out to $z\sim1$.  In Figure \ref{fig:tirlfevolution}, the TIR LFs are compared with literature values in the higher redshift bins. As shown in this figure, our TIR LF shown is broadly consistent with the published results in the redshift bins $[0-0.2]$, $[0.2-0.4]$ and $[0.4 - 0.8]$.  We find excellent agreement when we compare our {\vmax} LF in the redshift bins $[0.3-0.45]$ and $[0.6-0.8]$ to the values of \cite{lefloch2005632169} and \cite{ rodighiero20105158} in the same redshift bins. In the highest redshift bin, $[0.8-1.2]$, our TIR LF shows a good agreement with the LF determined by \cite{lefloch2005632169}. The SWIRE TIR LF however shows differences with the values of \cite{caputi200766097, magnelli200949657} and \cite{rodighiero20105158}. \citet{magnelli200949657} investigated the difference between their and the \cite{lefloch2005632169} LF values and found that the difference appears to be because of the choice of SED library and the correlation used to convert {$L_\lambda$} to {\lir}. Thus, this may also be the reason for the difference seen in the comparison of this work with theirs. Therefore although the {\vmax} LF values in the highest redshift bins are less robust, the parametric TIR LF and its evolution conforms to results that have been determined previously.

\begin{figure}
\begin{center}
\vspace{0.5cm}
\hspace{0.5cm}
\includegraphics[height=6.5cm, width=7cm]{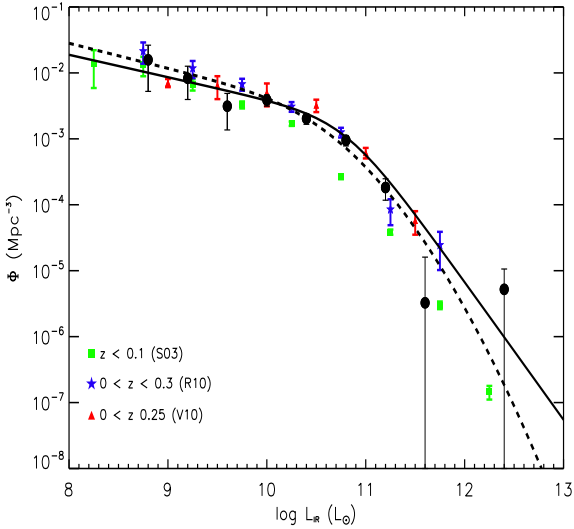}
\end{center}
\caption{The local ($0<z<0.2$) TIR LF determined using {\vmax} (filled black circles) and the parametric Bayesian method (solid and dashed black line) evaluated at $z=0.1$. The filled green squares are obtained from \citet{sanders20031261607} (S03), the filled blue stars are from \citet{rodighiero20105158} (R10) and the filled red triangles are from \citet{vaccari201051820} (V10).}
\label{fig:tirllf}
\end{figure}

\indent In Figure \ref{fig:tirllf}, we compare our local TIR LF with the \textit{IRAS} revised bright galaxy sample (BGS) derived local LF of \cite{sanders20031261607} (green filled squares) at $z<0.1$, {\spitzer} 24 {\micron} derived LLF of \cite{rodighiero20105158} (filled blue stars) at $z<0.3$ and HerMES SPIRE derived LLF of \cite{vaccari201051820} (filled red triangles) at $z<0.25$. The shape of the SWIRE local TIR LF and literature values are almost identical while the small difference between the local TIR LFs is most likely due to cosmic variance or the choice of the SED library used to calculate {\lir}. Several authors \citep{lefloch2005632169, rodighiero20105158, vaccari201051820} have reported best fit parameters of the local TIR LF by fitting to the binned LF for the power-law and log-normal LF model of Equation \ref{eq:saunders90lfmodel} using a $\chi^2$ minimisation procedure. The slope of the faint end is not well constrained and is usually fixed to the local value of $\sim1.2$, while $\sigma$ values in the range 0.39--0.72 have been reported. The {\loglstar} ({\lsun}) range is $\sim 9.24$--$10.6$ and {\logphistar} ({\mpcdex}) range is $\sim-2.00$ to $-2.06$ \citep{lefloch2005632169, rodighiero20105158, vaccari201051820}. The best fit parameters of our LLF for model 2 are \{{\loglstar} ({\lsun}), $\alpha, \sigma$, {\logphistar} ({\mpcdex})\} = \{9.87, 1.38, 0.68, $-2.26$\}, which are consistent with the published values.  For the double power-law model, \cite{sanders20031261607} find best-fit power-law indices $\alpha=-0.6\pm0.1$ and $\beta=-2.2\pm0.1$ and {\loglstar} $\sim10.5$ which are at least within the 99.7\% errors of the parameters estimated in our study.

\indent In Figure \ref{fig:tirlfevolution}, we compare our TIR LF with literature values in the higher redshift bins. As shown in this figure, our TIR LF shown is broadly consistent with the published results in the redshift bins $[0.2-0.4]$ and $[0.4 - 0.8]$. We find excellent agreement when we compare our {\vmax} LF in the redshift bins $[0.3-0.45]$ and $[0.6-0.8]$ to the values of \cite{lefloch2005632169} and \cite{ rodighiero20105158} in the same redshift bins. In the highest redshift bin considered here, our {\vmax} and the parametric LF show poor agreement because we are limited by the number of objects in these redshift bins. However when we compare the LFs with the values from \cite{caputi200766097}, and \cite{lefloch2005632169}, we find good agreement between the parametric LF and the literature values. Comparing our LF with the \cite{magnelli200949657} and \cite{rodighiero20105158} LF values (orange filled upside down triangles) shows a slight difference in that the binned estimates are lower than our parametric and {\vmax} LF. \citet{magnelli200949657} investigated the difference between their and the \cite{lefloch2005632169} LF values and found that the difference appears to be because of the choice of SED library and the correlation used to convert {\nulnu} to {\lir}. Thus, this may also be the reason for the difference seen in the comparison of our work with theirs. Therefore although our {\vmax} LF values in the highest redshift bins are less robust, our parametric TIR LF and its evolution conforms to results that have been determined previously.

\begin{table*}
\begin{center}
\begin{tabular}{ccccc}
\hline\hline
                                                                      &			 \multicolumn{4}{c}{$\log \Phi$ (Mpc$^{-3}$)}\\
$\log \mathrm{L_{IR}}$ (L$_\odot)$ 	     &           $0<{z}\leq0.2$             &      $0.2<{z}\leq0.4$              &       $0.4<{z}\leq0.8$            & $0.8<{z}\leq1.2$ \\
\hline
8.8 							     & $-1.80^{+0.22}_{-0.49}$ & 				        & 					&\\
9.2  							     & $-2.08^{+0.18}_{-0.32}$ & 				        & 					&\\
9.6   							     & $-2.51^{+0.19}_{-0.36}$ & 				        & 					&\\
10.0							     & $-2.42^{+0.08}_{-0.10}$ &					        & 					&\\
10.4 							     & $-2.69^{+0.07}_{-0.09}$ & $-2.96^{+0.24}_{-0.64}$ & 				         &\\
10.8 							     & $-3.02^{+0.07}_{-0.09}$ & $-2.73^{+0.11}_{-0.15}$ & 					&\\
11.2 							     & $-3.74^{+0.13}_{-0.19}$ & $-3.09^{+0.06}_{-0.06}$ & $-3.60^{+0.13}_{-0.19}$ &\\
11.6 							     & $-5.49^{+0.69}_{-5.49}$ & $-3.86^{+0.08}_{-0.10}$ & $-3.43^{+0.14}_{-0.21}$ &\\
12.0 							     &  --                                         & $-4.74^{+0.17}_{-0.30}$ & $-3.96^{+0.08}_{-0.09}$ & $-4.53^{+0.24}_{-0.56}$\\
12.4 							     & $-5.28^{+0.31}_{-5.28}$ & 				        & $-4.95^{+0.13}_{-0.18}$ & --\\
12.8 							     &					      & 				        & $-5.56^{+0.20}_{-0.38}$ & $-4.57^{+0.27}_{-0.88}$\\
13.2 							     & 					      & 				        & 				         & $-5.42^{+0.90}_{-5.42}$\\
\hline\hline
\end{tabular}
\end{center}
\caption{The {\vmax} TIR LF values.}
\label{tab:70mulfspecztirlfvmaxvalues}
\end{table*}

\section{Discussion}
\subsection{Evolution of the Infrared Luminosity Density}
\label{sect:cmirld}

Having determined the global evolution of the TIR LF in the redshift range $0<z<1.2$, we can now estimate how the comoving IR luminosity density (IRLD) and the space density of LIRGs and ULIRGs evolves with redshift. This is important because it allows us to evaluate the relative importance of IR luminous galaxies and their contribution to the CSFRD (associated with obscured star formation) out to $z\sim1.2$. 

\begin{figure}
\begin{center}
\vspace{0.5cm}
\includegraphics[height=6cm, width=7cm]{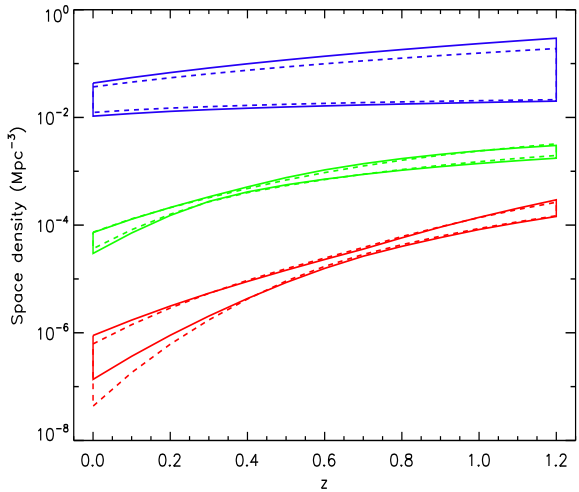}
\end{center}
\caption{Evolution of the space density of normal galaxies (solid and dashed blue region), LIRGs (solid and dashed green region) and ULIRGs (solid and dashed red region) up to $z=1.2$. The solid and dashed line regions includes 68\% of the posterior probability for the double power-law and power-law and log-normal LF models respectively.}
\label{fig:nz(u)lirgs}
\end{figure}

\indent We show in Figure \ref{fig:nz(u)lirgs}, the space density of normal galaxies (blue regions), LIRGs (green region) and ULIRGs (red region), which shows rapid evolution in the number density of LIRGs and ULIRGs while the number density of normal galaxies remains constant. The number density of LIRGs increases by a factor of $\sim43^{+11}_{-12}$ for the double power-law LF model and $\sim37^{+18}_{-06}$ for the power-law and log-normal LF model between $z=0$ and $z=1$, which is in agreement with the estimate of \cite{magnelli200949657}, who find a factor of 40. 

\indent \cite{magnelli200949657} estimate an increase in the number density of ULIRGs by a factor of $\sim100$ at $z\sim1$ than in the local Universe, while in Figure \ref{fig:nz(u)lirgs} ULIRGs increase by a factor of $\sim316^{+246}_{-138}$ for model 1 and for model 2 the factor is $\sim870^{+479}_{-499}$.  The difference between the values of model 1 and 2 is because of the steeper drop-off at the bright end of the LF for the power-law and log-normal LF model. Nonetheless, the 68\% credible intervals shown in Figure \ref{fig:nz(u)lirgs} for the two LF models are almost identical except for the ULIRG population, where the error for the power-law and log-normal LF model is larger. 

\indent We model the evolution of the space density of ULIRGs as $\propto(1+z)^n$, with $n=8.30^{+0.83}_{-0.83}$, which is consistent with the results of \cite{kim199811941} who find $n=7.6\pm3.2$. \cite{goto20114141903} use FIR data from \textit{AKARI} to find that ULIRGs evolve with $n=10.0\pm0.5$, which is similar to the value determined in our study. In contrast, \cite{jacobs2011141110} have estimated $n\simeq6\pm1$ using a sample of 160 {\micron} selected sources from {\spitzer} observations of the 1 {\sqrdeg} \textit{ISO} Deep Field region in the LH. Their analysis however is based on only 40 galaxies while the current study includes 634 sources and therefore the value determined here should be more accurate and reliable. 

\begin{figure*}
\begin{center}
\vspace{0.5cm}
\includegraphics[scale=0.55]{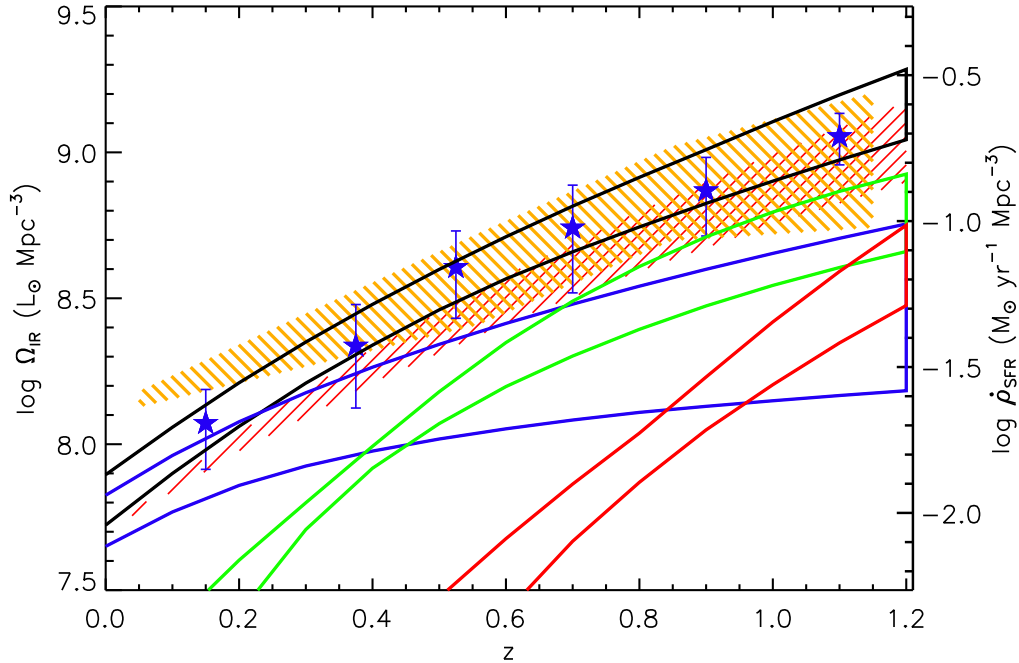}
\end{center}
\caption{Evolution of the IRLD up to $z=1.2$ (solid black line region) and the contribution by normal galaxies (solid blue line region), LIRGs (solid green line region) and ULIRGs (solid red line region) determined using the double power-law LF model. The black solid region includes 68\% of the posterior probability. The red line region is taken from \citet{lefloch2005632169}, orange line region is from \citet{magnelli200949657} and blue filled stars are taken from \citet{rodighiero20105158}.}
\label{fig:omegaIR}
\end{figure*}

\indent In Figure \ref{fig:omegaIR} we show the IRLD which is calculated by {\omegair} = $\int{L}\Phi(L, z|\{\theta\})\thinspace{dL}$. We only display IRLD determined using the double power-law LF model because the 68\% credible interval is almost identical when IRLD is calculated using the power-law and log-normal LF model. We follow the method presented in \cite{lefloch2005632169} by determining IRLD for IR luminous galaxies (LIRGs and ULIRGs) and ``normal" galaxies as a function of redshift. The evolution of {\omegair} is also represented in terms of an IR equivalent SFR using the calibration given by Equation \ref{eq:sfrk98}. The figure shows that our results are in agreement with what has been predicted from previous studies, particularly with the results of \citet{rodighiero20105158}. At the present epoch, most ($\sim85^{+03}_{-13}\%$) of the star formation activity is taking place in normal galaxies which have low extinction. As we move to higher redshifts, LIRGs evolve rapidly and dominate the star formation activity beyond $z\sim0.55^{+0.10}_{-0.05}$. 

\indent The model also suggests that, LIRGs and ULIRGs are responsible for $\sim68^{+10}_{-07}$ per cent of the total IRLD at $z=1.2$. At $z=0$, LIRGs and ULIRGs are responsible for less than $10^{+08}_{-02}\%$ of the IRLD, while at $z=1$, they produce $\sim66^{+10}_{-05}\%$ of it. These values are consistent with estimates of \cite{lefloch2005632169} while \cite{magnelli200949657} suggest that LIRGs and ULIRGs produce less than 2\% of the IRLD. This difference is most likely due to the fact that \cite{magnelli200949657} use the faint-end power-law index, $\alpha=-0.6$, from the {\iras} TIR LF of \cite{sanders20031261607}, which is steeper than the value used in this study ($\alpha=0.34^{+0.10}_{-0.12}$ for the double power-law LF model) and this would lead to a higher contribution of normal galaxies to the IRLD at $z=0$. 


\begin{figure*}
\begin{center}
\vspace{0.5cm}
\includegraphics[scale=0.55]{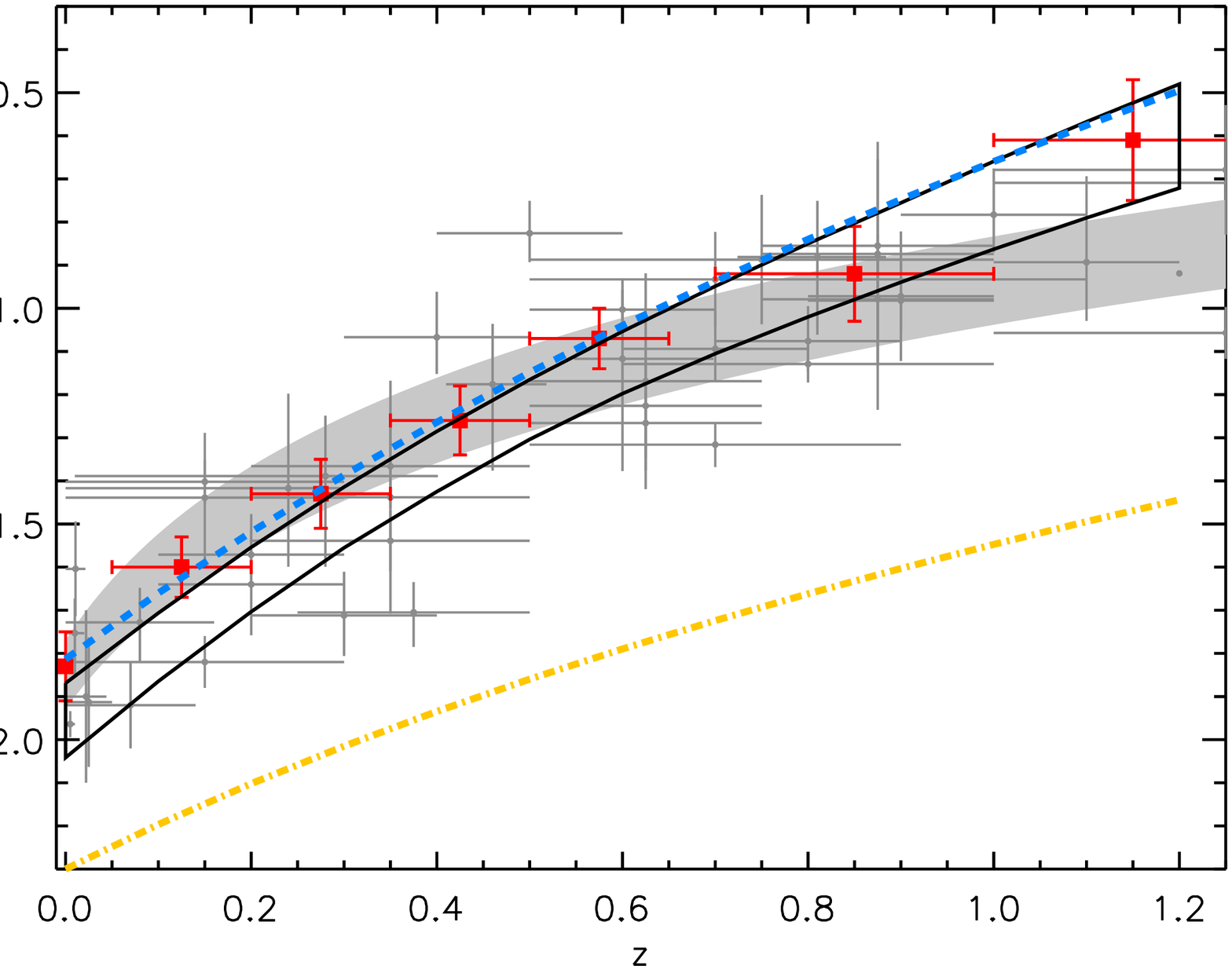}
\end{center}
\caption{Evolution of the CSFRD up to $z=1.2$ (solid black line region). The solid black line region includes 68\% of the posterior probability. The grey points are taken from \citet{hopkins2004615209} representing the CSFRD from various estimators. The red filled squares are CSFRD estimates from \citet{rujopakarn20107181171}. The orange dash-dot line is the UV dust extinction uncorrected CSFRD from \citet{schiminovich200561947} and the light blue dashed line represents the total CSFRD defined as the sum of the IR and UV dust extinction uncorrected SFR densities. The shaded grey region is the $3\sigma$ best-fitting regions CSFRD from \citet{hopkins2006651142}.}
\label{fig:cosmicsfr} 
\end{figure*}

\subsection{Evolution of the Cosmic Star Formation Rate Density}
\label{sect:csfrdensity}

\indent In Figure \ref{fig:cosmicsfr}, we show the CSFRD as a function of redshift up to $z=1.2$ and compare our results with the CSFRD estimates compiled by \cite{hopkins2004615209} and \cite{hopkins2006651142} at wavelengths ranging from the X-ray to radio. We highlight for comparison, in filled blue squares the SFR derived from 24 {\micron} observations taken from \cite{rujopakarn20107181171} (blue filled squares in Figure \ref{fig:cosmicsfr}). Overall, our prediction of {\rhosfr} show a good agreement with the results determined from previous experiments. At $z=0$, the CSFRD is $1.04^{+0.31}_{-0.13}\times10^{-2}$ {\sfrunits} which agrees with previous values of the local {\rhosfr} that have properly accounted for the internal extinction of the galaxy but slightly lower than the estimate of \cite{rujopakarn20107181171} who find a local {\rhosfr} of $1.65\times10^{-2}$ {\sfrunits} (assuming a \citealt{salpeter1955121161} IMF). The dust unobscured CSFRD estimated by \cite{schiminovich200561947} (dot dashed orange line) is also displayed in Figure \ref{fig:cosmicsfr}. At $z=0$, the dust unobscured CSFRD is estimated to be $\sim5.01\times10^{-3}$ {\sfrunits} and therefore without dust extinction corrections, the local CSFRD would be underestimated by $\sim65^{+8}_{-30}$ per cent. Modelling {\rhosfr} as $\propto(1+z)^Q$, implies $Q=3.82^{+0.25}_{-0.16}$, which is higher than the value reported by \cite{hopkins2004615209} ($Q=3.29\pm0.26$) but consistent with \cite{rujopakarn20107181171}  ($Q=3.50\pm0.20$).  The results presented here show that in general, most of the star formation over the last 8 billion years has taken place in dust obscured galaxies. 

\subsection{Future prospects}
\label{sect:swirelffutureprospects}
In Section \ref{sect:results}, we have presented the results of the evolution of the far-IR luminosity functions in the SWIRE XMM-LSS and Lockman Hole fields, based on a spectroscopic redshift sample. The 70 {\micron} and TIR LFs were constructed using the sample selected with S$_{70} >10$ mJy and $r<22$ and the 160 {\micron} LF was constructed using sources with S$_{160} >60$ mJy and $r<22$. In this section we discuss the limitations of the sample selection and the methodology in constructing the FIR LFs and suggest improvements that could be carried out in order to obtain better estimates of LFs.

\indent We chose to study the evolution of the FIR LFs using only spectroscopic redshifts in order to reduce the uncertainties related to photometric redshifts as stated in Section \ref{sect:introswirelf}. Some previous studies of the evolution of the IR LFs that have utilised photometric redshift \citep{perezgonzalez200563082, babbedge20043701159,wang201040135} account for the uncertainty in the redshift by performing Monte Carlo analysis, where each source's photometric redshift is replaced by a redshift drawn randomly from a Gaussian distribution centred on the original photometric redshift and then iterating the LF procedure.  Few authors \citep{chen2003586745, aird20104012531} have attempted to account for this uncertainty by modifying the likelihood function in Equation \ref{eq:lhfunc} and including the actual redshift probability distribution function for each source likelihood ($p(\{L_i, z_i\}|\{\theta\})$) rather than a gaussian approximation. Therefore including sources with photometric redshifts will not only increase the sample size but we will be able to apply a fainter magnitude cut in the optical and reach higher redshifts. 

\indent At IR wavelengths, the sample includes sources up to the flux limit of the SWIRE survey but with a strict magnitude cut of $r<22$ our sample will not include those IR sources that are also optically faint. For the present sample, $>50\%$ of the 70 {\micron} sources at 10 mJy have an optical counterpart that is brighter than 22 mag in the $r-$band and therefore inclusion of the photometric redshifts will also reduce the dependence on the spectroscopic incompleteness weights $w_i$. Since future large surveys will discover many millions of galaxies, photometric redshifts will play a vital role in any statistical analysis and therefore the inclusion of this in our studies will be one of the next key step.

\indent  The other limitation of the work is that in the parametric method, we have assumed a single evolving population. For the 70 {\micron} and the TIR LF the selection functions are calculated assuming an M82 SED template and for the 160 {\micron} LF we have used a mixture of M82 and Cirrus SED. As shown in Figure \ref{fig:lirvsz}, we observe a gradual change in the best-fit SED type as we move out to higher redshift, with low redshift sources fit with a cirrus dominated SED and the higher redshift sources fit with a starburst dominated SED. This implies that the starburst component of the IR SED must evolve much faster than the cirrus component and therefore this evolution must be taken into account when studying the evolution of LFs using parametric methods. Therefore studying the evolution of each SED component will be the second step in studying the evolution of the FIR LFs. 

\section{Summary and Conclusion}
\label{sect:swirelfsummary}
We have presented a new observational determination of the FIR LFs using 70 {\micron} selected sources from the SWIRE survey with spectroscopic redshifts. The primary sample was selected from the photometric redshift catalogueue of \cite{rowanrobinson2008386687}, which contains over 1 million IR sources estimated by combining optical and IRAC 3.6 and 4.5 {\micron} photometry. We computed the rest-frame 70 {\micron} and TIR LF using sources with $S_{70}>10$mJy and $r<22$ and studied their evolution out to $z=1.2$. The evolution of the 160 {\micron} LF was determined using $S_{160}>60$mJy and $r<22$ out to $z=0.5$.

\indent We use the multiwavelength optical to IR data to model the SED for each source in our sample to estimate the rest-frame monochromatic luminosities to determine the 70 and 160 {\micron} LFs using Bayesian parametric and the {\vmax} methods. The work presented here is an improvement on earlier works because we use 70 {\micron} data rather than rely on extrapolations from 24 {\micron}. In addition, the Bayesian method is a new approach to FIR astronomy, which can be used to further the study of IR LFs. For example we can include photometric redshift data and properly account for the individual photometric redshift uncertainty of each source to investigate the evolution of the FIR LFs. The study of the evolution of FIR LFs using photometric redshifts will be presented in a future paper. In our analysis we have corrected for the optical and IR selection biases and incompleteness of the spectroscopic sample, which have allowed us to accurately construct the LFs. 

\indent The parametric Bayesian FIR LFs were determined using two LF models; a double power-law and a power-law and log-normal models. Comparison of the two models with the binned {\vmax} estimates showed that both LF models provide a good fit to the 70 {\micron} and TIR LFs. For the 160 {\micron} LF however, the power-law and log-normal LF was found to underestimate the number density of sources at {\lonesixty} $>10^{11}$ {\lsun} when compared to the binned LF estimates, implying that the double power-law LF model provides a better description of FIR LFs. 

\indent The evolution of FIR LFs was modelled using a pure luminosity evolution model. The rest-frame 70 {\micron} LF was found to evolve rapidly in luminosity with {\alphal} $=3.39^{+0.12}_{-0.22}$ for model 1 and {\alphal} $=3.41^{+0.18}_{-0.25}$ for model 2. Thus the choice of LF model does not have a large effect on the evolution of the 70 {\micron} LF. The 160 {\micron} LF was found to evolve with {\alphal} $=5.73^{+0.29}_{-0.45}$ for model 1 and {\alphal} $=5.62^{+0.21}_{-0.45}$ for model 2. The faster rate of evolution of the 160 {\micron} LF is consistent with the study of \cite{takeuchi2006448525} which finds the \textit{ISO} 170 {\micron} LF evolves with {\alphal} $= 5.0^{+2.3}_{-0.5}$. Since cooler galaxies are detected more effectively at 160 {\micron} than at shorter wavelengths, the strong evolution implies the presence of large amounts of cool dust at higher redshifts.

\indent Finally, we used the best-fit SEDs to estimate total IR luminosities of each source to derive the TIR LF to $z=1.2$. The SWIRE TIR LF showed good agreement when compared with literature values estimated from {\spitzer} and {\herschel} data across all redshifts bins. The TIR LF evolves with $\text{\alphal}=3.82^{+0.23}_{-0.20}$ for model 1 and $\text{\alphal}=3.82^{+0.25}_{-0.16}$ for model 2 corroborating the conclusions of previous results from {\spitzer} 24 {\micron} studies which find strong luminosity evolution. The TIR LF was then integrated to calculate the IRLD out to $z=1.2$, which confirms the rapid evolution in number density of LIRGs which contribute $\sim66^{+10}_{-5}$ per cent to the IRLD, and hence the CSFRD at $z=1$. The results presented in this chapter, based on 70 {\micron} data confirm that the bulk of the star formation at $z=1$ takes place in dust obscured objects.

\section{Acknowledgements}
HP acknowledges financial support from STFC. This work is based on observations made with the {\spitzer} Space Telescope, which is operated for NASA by the Jet Propulsion Laboratory, California Institute of Technology.

\bibliographystyle{mn2e}
\bibliography{swire_lfs}
\nocite{*}

\end{document}